\documentclass[12pt,a4paper,openany]{book}
\usepackage{graphicx}
\usepackage{setspace}	
\usepackage[numbers]{natbib}
\usepackage[utf8]{inputenc}
\usepackage{color}
\usepackage{hhline} 		
\usepackage{multirow}
\usepackage{subfigure}
\usepackage{acronym}
\usepackage[hidelinks]{hyperref}
\usepackage{amsmath,amsmath,amssymb}
\usepackage{fancyhdr}
\usepackage{epsfig, amsmath}
\usepackage[ruled,vlined]{algorithm2e}
\definecolor{commentcolor}{gray}{0.4}

\SetCommentSty{mycommfont}
\usepackage[acronym]{glossaries}
\usepackage{amsthm}
\usepackage{csvsimple}
\usepackage{booktabs}
\usepackage{cleveref}
\usepackage[toc]{appendix}
\usepackage[pdf]{graphviz}

\theoremstyle{definition}
\newtheorem{definition}{Definition}[section]

\hypersetup{
	linktocpage=true,
	colorlinks=true,
	linkcolor=[rgb]{0,0.2,0.2},
	citecolor=[rgb]{0.2,0,0.2},
	urlcolor=[rgb]{0.5,0.1,0},
}

\newacronym{rl}{RL}{Reinforcement Learning}
\newacronym{recsys}{RS}{Recommender System}
\newacronym{ltr}{LTR}{Learning To Rank}
\newacronym{ctr}{CTR}{Click-through Rate}
\newacronym{dl}{DL}{Deep Learning}
\newacronym{drl}{DRL}{Deep Reinforcement Learning}
\newacronym{mdp}{MDP}{Markov Decision Process}
\newacronym{dqn}{DQN}{Deep Q-Networks}
\newacronym{ddqn}{DDQN}{Double Deep Q-Networks}
\newacronym{rnn}{RNN}{Recurrent Neural Network}
\newacronym{a2c}{A2C}{Advantage Actor-Critic Networks}
\newacronym{srs}{SRS}{Sequential Recommender System}
\newacronym{ndcg}{NDCG}{Normalized Discounted Cumulative Gain}
\newacronym{ucb}{UCB}{Upper Confidence Bound}
\newacronym{ltv}{LTV}{Life-Time Value}
\newacronym{erb}{ERB}{Experience Replay Buffer}
\newacronym{per}{PER}{Prioritized Experience Replay}
\newacronym{iid}{i.i.d.}{independent and identically distributed}
\newacronym{dp}{DP}{Dynamic Programming}
\newacronym{td}{TD}{Temporal-Difference}
\newacronym{relu}{ReLU}{Rectifier Linear Unit}
\newacronym{mse}{MSE}{Mean Square Error}
\newacronym{pomdp}{POMDP}{Partially Observable Markov Decision Processes}
\newacronym{ml}{ML}{Machine Learning}
\newacronym{ui}{UI}{User Interface}
\newacronym{ann}{ANN}{Artificial Neural Networks}
\newacronym{svd}{SVD}{Singular Value Decomposition}
\newacronym{trpo}{TRPO}{Trust Region Policy Optimization}

\newglossaryentry{coldstartproblem}
{
    name=Cold-Start Problem,
    description={is a problem of recommendation systems that emerges when recommendations need to be made for a user who we don't have enough data about}
}
\newglossaryentry{collaborativefiltering}
{
    name=Collaborative Filtering,
    description={in recommender systems is the concept of using one user's preferences to infer other's}
}

\newglossaryentry{knowledgebased}
{
    name=Knowledge-Based,
    description={are recommendation systems that rely on user created constraints, and domain-specific product matching rules }
}

\newglossaryentry{casebased}
{
    name=Case-Based,
    description={are knowledge-based systems that let users provide a sample item, and recommend similar ones}
}

\newglossaryentry{longtermpredictions}
{
    name=Long-Term Prediction,
    description={are predictions that will be eventually consumed by a user. In contrast, short-term predictions aim to be immediately consumed}
}
\newglossaryentry{interactiverecommendersystems}
{
    name=Interactive Recommender Systems,
    description={are recommendation systems that receive direct user feedback on the relevancy of recommendations}
}
\newglossaryentry{supervisedlearning}
{
    name=Supervised Learning,
    description={are a class of machine learning algorithms that address regression and classification problems}
}

\makeglossaries

\begin{document}

\frontmatter

\newpage
\thispagestyle{empty}

\baselineskip 2em


\centerline{\includegraphics[width=0.6\textwidth]{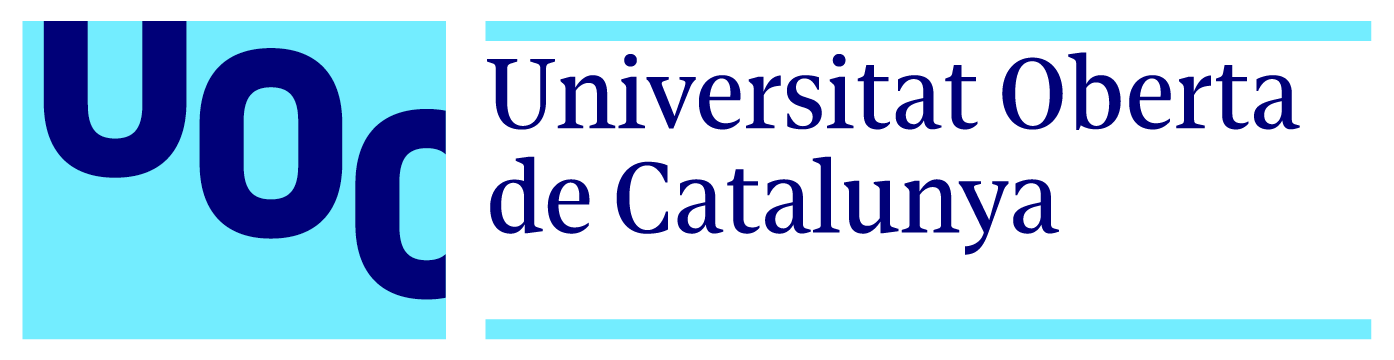}}
\begin{center}
\textsc{Universitat Oberta de Catalunya (UOC) \\
 Máster Universitario en Ciencia de Datos (\textit{Data Science})\\}


\vspace*{1.5cm}

\textsc{\Large TRABAJO FINAL DE MÁSTER}

\vspace*{0.5cm}

\textsc{\large Área: 2}


\vspace*{2.0cm}

\textbf{\Large Optimized Recommender Systems with Deep Reinforcement Learning}

\vspace{2.5cm}
\baselineskip 1em

\baselineskip 2em
-----------------------------------------------------------------------------\\
Autor:      Lucas Farris\\
Tutor:      Luis Esteve Elfau\\
Profesor:   Jordi Casas Roma\\
-----------------------------------------------------------------------------\\
\vspace*{1.5cm}
Barcelona, \today

\end{center}

\newpage
\pagestyle{empty}
\hfill

\newpage
 \setcounter{page}{1} 
\pagestyle{plain}

\chapter*{Créditos/Copyright}

\vspace{1cm}

\begin{figure}[ht]
    \centering
	\includegraphics[scale=1]{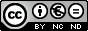}
\end{figure}

Esta obra está sujeta a una licencia de Reconocimiento -  NoComercial - SinObraDerivada

\href{https://creativecommons.org/licenses/by-nc-nd/3.0/es/}{3.0 España de CreativeCommons}.

\chapter*{FICHA DEL TRABAJO FINAL}

\begin{table}[ht]
	\centering{}
	\renewcommand{\arraystretch}{2}
	\begin{tabular}{r | l}
		\hline
		Título del trabajo: & \begin{tabular}{@{}c@{}}Optimized e-commerce recommendations with \\ deep reinforcement learning\end{tabular} \\
		\hline
        Nombre del autor: & Lucas Farris\\
		\hline
        Nombre del colaborador/a docente: & Luis Esteve Elfau\\
		\hline
        Nombre del PRA: & Jordi Casas Roma \\
		\hline
        Fecha de entrega (mm/aaaa): & \today \\
		\hline
        Titulación o programa: &  Máster Universitario en Ciencia de Datos\\
		\hline
        Área del Trabajo Final: & M2.879 - Área 2 \\
		\hline
        Idioma del trabajo: & Inglés\\
		\hline
		Palabras clave & \begin{tabular}{@{}c@{}}deep reinforcement learning, recommender systems, \\ e-commerce recommendations\end{tabular} \\
		\hline
	\end{tabular}
\end{table}

\chapter*{Dedication}

This is for my aunt, Maria Giuliana Farris, may she rest in peace.
She helped me get the computer, and courage, with which these words were typed.


\chapter*{Acknowledgements}

I would like to thank my advisor, Luis Esteve Elfau, for his guidance, trust, and for helping me prioritize my time in this work. Also, to thank the \textit{Universitat Oberta de Catalunya} and its staff for the opportunity. 

Finally I would like to thank my brother Pedro, my mother Lilian, and my beloved Natalia, for supporting me throughout the years.

\chapter*{Abstract}
\addcontentsline{toc}{chapter}{Abstract}

\onehalfspacing

Recommender Systems have been the cornerstone of online retailers. Traditionally they were based on rules, relevance scores, ranking algorithms, and supervised learning algorithms, but now it is feasible to use reinforcement learning algorithms to generate meaningful recommendations. This work investigates and develops means to setup a reproducible testbed, and evaluate different state of the art algorithms in a realistic environment. It entails a proposal, literature review, methodology, results, and comments.

\section*{\Huge Resumen}
El Sistema de Recomendaciones es parte importante de un comercio electrónico. Tradicionalmente estos sistemas se basaban en algoritmos de reglas, métricas de relevancia, clasificaciones y aprendizaje supervisado. Actualmente se pueden usar algoritmos de aprendizaje por refuerzo para crear recomendaciones más útiles. Este trabajo investiga y desarrolla maneras de producir un sistema de prueba de agentes, y comparar diferentes algoritmos de última generación en un entorno de simulación basado en datos reales. Se describe una propuesta, una revisión de la literatura, la metodología de investigación, los resultados obtenidos y comentarios.

\vspace{1.5cm}

\textbf{Keywords}: deep reinforcement learning, recommender systems, e-commerce recommendations

\newpage

\pagestyle{fancy}
\renewcommand{\chaptermark}[1]{ \markboth{#1}{}}
\renewcommand{\sectionmark}[1]{\markright{ \thesection.\ #1}}
\lhead[\fancyplain{}{\bfseries\thepage}]{\fancyplain{}{\bfseries\rightmark}}
\rhead[\fancyplain{}{\bfseries\leftmark}]{\fancyplain{}{\bfseries\thepage}}
\cfoot{}

\clearpage
\phantomsection
\addcontentsline{toc}{chapter}{Index}
\tableofcontents
\clearpage
\phantomsection
\addcontentsline{toc}{chapter}{List of Figures}
\listoffigures
\clearpage
\phantomsection
\addcontentsline{toc}{chapter}{List of Tables}
\listoftables
\clearpage
\phantomsection
\addcontentsline{toc}{chapter}{Glossary}
\printglossary
\addcontentsline{toc}{chapter}{Acronyms}
\printglossary[type=\acronymtype]

\let\cleardoublepage\clearpage

\mainmatter

\pagestyle{fancy}
\renewcommand{\chaptermark}[1]{ \markboth{#1}{}}
\renewcommand{\sectionmark}[1]{\markright{ \thesection.\ #1}}
\lhead[\fancyplain{}{\bfseries\thepage}]{\fancyplain{}{\bfseries\rightmark}}
\rhead[\fancyplain{}{\bfseries\leftmark}]{\fancyplain{}{\bfseries\thepage}}
\cfoot{}

\onehalfspacing

\chapter{Introduction}

\label{chapter:introduccion}

\section{Topic Relevance}

Due to recent developments in large scale industrial production, improvements in logistics, and scalable technological infrastructure, e-commerce platforms are now able to offer thousands of products to its users. One of the challenges that arise in this scenario is how to properly recommend products, given such large variety but little space for recommendations. A solution that addresses this problem is commonly known as a \acrfull{recsys}.

Search engines like \textit{Apache Lucene} rely on users typing or selecting filters that approximately match what they are looking for. They usually calculate \textit{relevance scores} that estimate how well each document matches each search term, and order results based on it \cite[see][chap.~3]{lucene}. Some \acrshort{recsys} leverage features such as user-made evaluations, search history, demographic data, and geographic data. Then, they may apply a family of supervised learning ranking algorithms like Learning-To-Rank (LTR) \citep[see][chap.~13]{falk_practical_2019}. Such systems may even estimate which products have the highest probability of being chosen by the end user.

Alternatively, there are methods based on \acrfull{rl} that may take into account a user's journey. They make recommendations not only based on search filters or user data, but also on a user's interaction and behavior (i.e., what buttons were clicked and when). The usual goal of such agents is to optimize for metrics like \acrfull{ctr}, conversion rate, or dropout rate. In other words, to increase the percentage of users that are clicking on products or buying them, or to decrease the percentage of users that leave without selecting recommendations. 

The proposal for this work is to leverage interaction data from large retailers, use them to generate a \acrshort{rl} environment, and measure how different \acrfull{drl} algorithms perform under these circumstances. There has been research on ways generate faithful user simulations from website interaction data, and how to use such simulations to create realistic learning environments. The goal is not to improve the simulations or environments, but rather perform a comprehensive evaluation on which methods work best.

The relevance of this project can be understood from three points of view. There is the scientific relevance of applying the latest advancements in \acrshort{drl} to new problems and publish the results. There is the social benefit of developing technologies that help humans find what they need more easily - especially in the information era of overwhelming amounts of data. Lastly, there is the commercial relevance, since online shopping has been on the rise, and retailers often struggle to help consumers find what they need. 

\section{Personal Motivation}

Personally, as many other online consumers, I have had trouble finding products online. More often than not I knew almost exactly what I was looking for but couldn't find relevant products in the recommendations. In many cases it was only after exhaustive searching through pages of recommendations that I found what I was looking for. 

From a professional point of view, I'm also very intrigued by this problem. In 2018 I started working as an engineer in an \textit{Elasticsearch} partner firm. I've had the chance to study and learn how modern search engines estimate relevance and can scale to millions of users. In 2019 I've worked on improving marketing campaigns by predicting psychological profiles from users, based on their data. In 2020 I've had the opportunity to work on a solution based on \acrshort{rl} that used interaction data to dynamically customize the user interface, with the goal of improving engagement. 

Another personal motivation for me was the book \textit{Persuasion Profiling} by \citet{persuasion}. It explains, among other things, why online stores fail to reach the conversion rates of their physical counterparts. The author describes this issue from a decision making standpoint, and discusses concepts that could be implemented to improve the performance of online retailers. I believe, that there is a lot of potential in modeling the recommendation problems as a \acrfull{mdp}, and solving it using \acrshort{drl}.

\section{Main and Secondary Goals}

The main goals of this work are to use an existing \acrshort{rl} environment (containing an interaction dataset and a user simulator), apply state of the art DRL algorithms to it, thoroughly compare different algorithms, and evaluate which ones work best under which conditions.

In terms the environments, algorithms, and metrics, we are still evaluating which are more relevant in the literature. Currently the best potential environment we found was \textit{MARS-gym} \citep{mars}, which is based data from the \textit{ACM Trivago RecSys 2019 Challenge} \citep{challenge}. A few potential algorithms are: Deep Q-Learning (DQN), Double Deep Q-Learning (DDQN), Categorical DQN, Rainbow DQN, REINFORCE, and Advantage Actor-Critic (A2C). For the metrics we would like to compare standard RL metrics such as the cumulative mean reward, and more traditional RecSys metrics like catalog coverage \cite{recsysmetrics}. This metric measures what percentage of the product catalog was displayed to a user during a session.

The partial goals for the project are setting up the user simulations, setting up the environments, implementing each of the RL algorithms, and creating a testbed for reproducible comparison. If the main goals are reached in sufficient time, it would be interesting to evaluate how trained RL models perform after perturbations in the user simulator (to simulate seasonality in e-commerce websites). Another potential direction of investigation would be to evaluate if a model, pre-trained in a certain environment, performs well in a new but similar environment (to simulate adaptation to new demographics). 

\section{Methodology}

Popular methodologies such as CRISP-DM \cite{crispdm} focus on the development of commercial and stable products. Therefore, it would be pragmatic to use a simplified version of them in an investigation effort. We propose a reduced version of CRISP-DM, without the \textit{Business Understanding} and \textit{Deploy} steps. A sequential diagram of this methodology can be found in \cref{fig:crisp}. The scope of the remaining steps would be:

\begin{figure}[!htbp]
	\centering
	\includegraphics[width=0.8\textwidth]{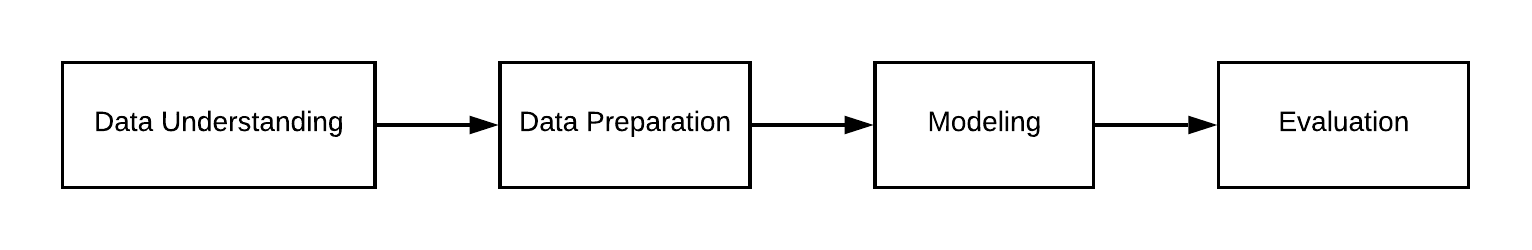}
	\caption{Adaptation of CRISP-DM focusing on investigation, rather than commercial application.}
	\label{fig:crisp}
\end{figure}

\begin{enumerate}
  \item \textbf{Data Understanding}: to obtain and understand the Trivago tracking data
  \item \textbf{Data Preparation}: to use the tracking data to generate a simulator, a RL environment, and a testbed
  \item \textbf{Modeling}: to implement several RL agents to be trained in the environment
  \item \textbf{Evaluation}: to run several variations of each agent, measure, and compare them.
\end{enumerate}

\section{Project Schedule}

\begin{figure}[!htbp]
	\centering
	\includegraphics[width=0.9\textwidth]{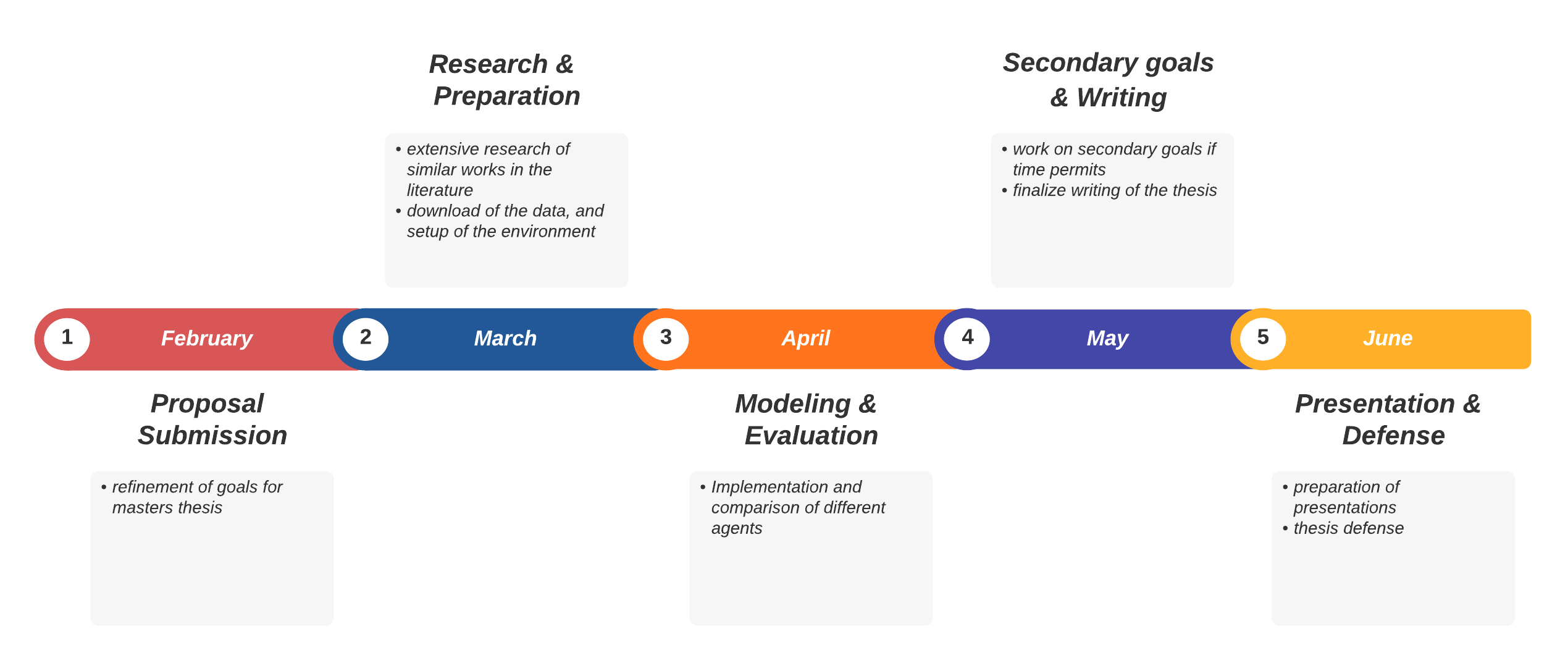}
	\caption{Timeline of development of the thesis.}
	\label{fig:timeline}
\end{figure}

The planned schedule for the development of the thesis is represented in \cref{fig:timeline}. It specifies the order and dependencies of the multiple efforts that need to take place for the work to be completed.

The month of February is dedicated to finalizing this proposal, by defining goals and refining what we want to achieve. Then, during the first three weeks of March (until the $20^{th}$), the focus will be in researching the state of the art of using \acrlong{ml} (\acrshort{ml}) in \acrshort{recsys}, and finding research related to the scope of this thesis. The most significant challenges for me in this step are to get familiar with the recommender system literature, and to find the best sources to review in the field.

Once the state of the art and related works are researched, we will concentrate on the implementation of the goals, until May $23^{rd}$. This phase will consist in setting up the source dataset, the user simulations, the environments, and the algorithms. It will also include the execution and measurement of all comparisons. I have experience implementing and evaluating DRL models and using Gym environments, but I lack experience with training recommender systems on offline data, and specially in building user simulations from tracking data. I'm also not very familiar with the metrics commonly used in the \acrshort{recsys} literature.

The next step of the project will be finalizing the writing of thesis, by appropriately documenting the methodology and findings. This step will also account for the preparation of presentations, and will finish on June $6^{th}$. Finally the last days of the project will contain the formal presentations, and defenses.

\chapter{Literature Review}

\label{chapter:sota}

\section{Investigation Method}

For the purpose of reviewing the literature on applying \acrshort{rl} to address the recommendation problem, and analyzing the state-of-the-art, the methodology described by \citet[chap.~4]{writingforscience} was used. The literature review will first cover a short historical overview of the fields of reinforcement learning and recommender systems. Secondly, based on some of the most influential textbooks off each field, a review of standard techniques will be provided. Then we'll explore a few recent review articles to have a panorama view of the field. Finally, the key papers of the area will be analyzed, along with some more recent fringe papers.

\section{Historical Background}

Recommender Systems grew in popularity during the 1990s. One of the first such systems was \textit{GroupLens} which was also responsible for publicly releasing large datasets that significantly helped researchers in the area \cite[chap.~1.2]{recommendersystems}. In more recent years, thanks to tech giants like \textit{Netflix} and \textit{Amazon} this field has gained a lot of attention, and one of its hallmarks is the annual ACM \textit{RecSys} challenge.

The Reinforcement Learning field as we know it came together in the 1980s, thanks to a combination of optimal control research (that started in the 1950s with Bellman) and psychological research of animal learning. A richly detailed history of the field can be found in \cite[chap.~1.7]{reinforcementlearning}.

Artificial neural networks research started in the 1940s with \textit{cybernetics} researchers attempting to generate computational models of the brain. Then in the 80s under the name \textit{connectionism} there was the development of \textit{back-propagation}, and \textit{multi-layer networks}. Deep Learning research started in 2006 \cite[chap.~1.2.1]{deeplearning}.

\section{Standard Definitions and Techniques}

Most technical information described in this section was summarized from influential textbooks in the fields of Recommender Systems \cite{recommendersystems}, Reinforcement Learning \cite{reinforcementlearning}, and Deep Learning \cite{deeplearning}.

\subsection{Recommender Systems}

The problem that Recommender Systems solve can be generalized as in \cref{def:recsysproblem}. Using this generic definition, we can explore all the types of recommender systems under a unified view.

\begin{definition}[The Recommender System Problem]
\label{def:recsysproblem}
To deliver a set of items, to a set of users, optimizing a set of goals.
\end{definition}

Some systems achieve this by predicting certain user-item metrics. Examples of metrics that could be predicted are click-throughs, user ratings, or relevance ranking. The most usual ones though are ratings of user-product tuples, and the top-k highest ranked items for a given user. Systems that use the latter are sometimes known as Top-N Recommender Systems. Common goals for such systems include accuracy, relevance, novelty, serendipity, and diversity.

The predicted items could be movies in a streaming platform, books in an online store, ads in a marketing platform, or friend connections in a social network. The entities that receive predictions are usually users of an online system, but could also be businesses or even software agents.

Metrics, items, and users can be modeled in different ways. For instance, product ratings could be measured as a like or dislike (\textit{binary}), a 5-star rating (\textit{numerical}), a Likert scale (\textit{categorical}), or even a button click (\textit{implicit}). Items could be wine brands, and be modeled by the features that make them unique, for instance acidity, sweetness, bitterness, aroma, color, and aftertaste. Users could be modeled using demographic data, geographic data, their purchase history, their user interactions, and so on.

One way of presenting these concepts is by using a \textit{Utility Matrix} \cite[see][chap.~9]{leskovec_mining_2019}. Some studies refer to the modeling of user-item goals as \textit{utility}, and this matrix representation places users in rows, items in columns, and measurements for the cells. The \cref{table:utilitymatrix} contains an example of this concept.

\begin{table}[!htbp]
\centering
\begin{tabular}{l}
\begin{filecontents*}{csv/utilitymatrix.csv}
,Item 1,Item 2, Item 3,...,Item M
User 1,1,0,0,...,2
User 2,0,2,0,...,1
...,...,...,...,...,...
User N,1,3,0,...,0
\end{filecontents*}
\csvautotabular{csv/utilitymatrix.csv}
\end{tabular}
\caption{Example of a $N \times M$ Utility Matrix. Cell values represent utility or relevance, for instance how many times the user at that row clicked on the product at that column. Note that in practice, this is often a very sparse matrix.}
\label{table:utilitymatrix}
\end{table}

\subsubsection{Recommender System Categorization}

If the main features of a recommendation system are based on user-item interactions, it is known as a \gls{collaborativefiltering} system. For example, a collaborative filtering wine recommendation system could recommend you a bottle because this particular brand has recently received high ratings from other users with similar interests as you. Such systems are challenging because user-interaction data is usually sparse. The two main methods of collaborative filtering are Memory-Based, and Model-Based, according to how predictions are made.

Memory-Based methods make predictions based on the neighborhood of the current user. In other words, a distance metric is chosen (e.g., Euclidean distance, Manhattan distance), and computed pairwise between users. These methods are again subdivided into two categories, according to how predictions are made. \textit{User-Based Collaborative Filtering} is a special case of memory-based methods that find products that likeminded users chose in the past. Meanwhile \textit{Item-Based Collaborative Filtering} finds items similar to what a particular user liked in the past. 

Model-Based methods, in contrast, use \acrshort{ml} to predict item metrics for users. It has been shown that combinations of Memory-Based and Model-Based systems can provide very accurate results. It has also been shown that \gls{supervisedlearning} can be generalized to effectively tackle the recommendation problem.

If the main features are modeled after product attributes, it is called a Content-Based system. In such systems user interaction is combined with item data into features, and the rating is predicted as the target of a \gls{supervisedlearning} algorithm. Such systems often provide more obvious recommendations (this issue is known as \textit{low serendipity}), and are not effective in predicting items for newer users - this is known as the \gls{coldstartproblem}. In contrast, they perform better when predicting ratings for new products (because many models are able to generalize knowledge based on product metadata).

Systems that are modeled after user-specified constraints (e.g., search fields, filters, ranges) are called \gls{knowledgebased} systems. They are usually applied when there's limited domain knowledge, when user ratings tend to change over time, or due to seasonality. Such systems usually contain business-specific rules for determining product similarities, and they allow users to explicitly query what they're looking for. These systems are labeled according to how users input requirements. Constraint-Based systems allow users to specify certain values or ranges for specific attributes of items (e.g., filtering real state by square meter price). In \gls{casebased} systems users provide sample items, or specifications, and it will find similar items (e.g., filtering real state by desired number of bedrooms). 

Depending on a \gls{knowledgebased} system's user interface, it can be categorized differently. In Conversational Systems, users are questioned about their needs via natural language dialogue (e.g., a chatbot), in Search-Based Systems users answer a preset number of questions (e.g., a quiz). In Navigation-Based Systems, also known as \textit{Critiquing-Based Recommender Systems}, users iteratively request changes to an example product (e.g., a car dealership website that allows users to find similar cars with lower carbon emission). Finally, a special case of \gls{knowledgebased} systems are called Utility-Based, for when the relevance (i.e., item utility) formula is known \textit{a priori}. 

\textit{Hybrid Recommender Systems} are the ones that combine aspects of some of the different techniques already mentioned. Other subcategories of recommender systems include: demographic recommender systems (when user demographic data is used), context-based or context-aware systems (when time, location, social data are used), time-sensitive systems (ratings evolve or depend on seasonality), location-based systems (user locations are used). Special types of recommender systems are social recommenders. These models recommend using social cues like social graph connections (i.e., using \textit{PageRank} algorithms), social influence (also known as viral marketing or influence analysis), trustworthiness (by having users explicitly provide trust/distrust ratings), and social tagging (e.g., hashtags).

Recommendation systems that model users in groups are known as \textit{Group Recommender Systems}. Systems that allow each item attribute to have its own rating (for instance hotel booking websites, that rate hotels on cleanliness or hospitality) are known as \textit{Multi-Criteria Recommender Systems}. The aspect of some recommendation systems to actively encourage users to rate less popular items is known as \textit{Active-Learning}. 

There are some less popular definitions that are not present in the textbook, but are part of the more recent literature. Some researches refer to systems that directly receive user feedback on recommendations as Interactive Recommender Systems. Some researchers refer to systems that actively leverage user interaction to make predictions as Sequential Recommender Systems. Finally, some make a distinction between how soon users are expected to accept recommendations. In this sense, Long-Term Prediction systems are optimized to have users select items at any time in the future. Meanwhile Short-Term Prediction Systems expect users to select recommendations immediately.

\begin{figure}[!htbp]
	\centering
	\includegraphics[width=1\textwidth]{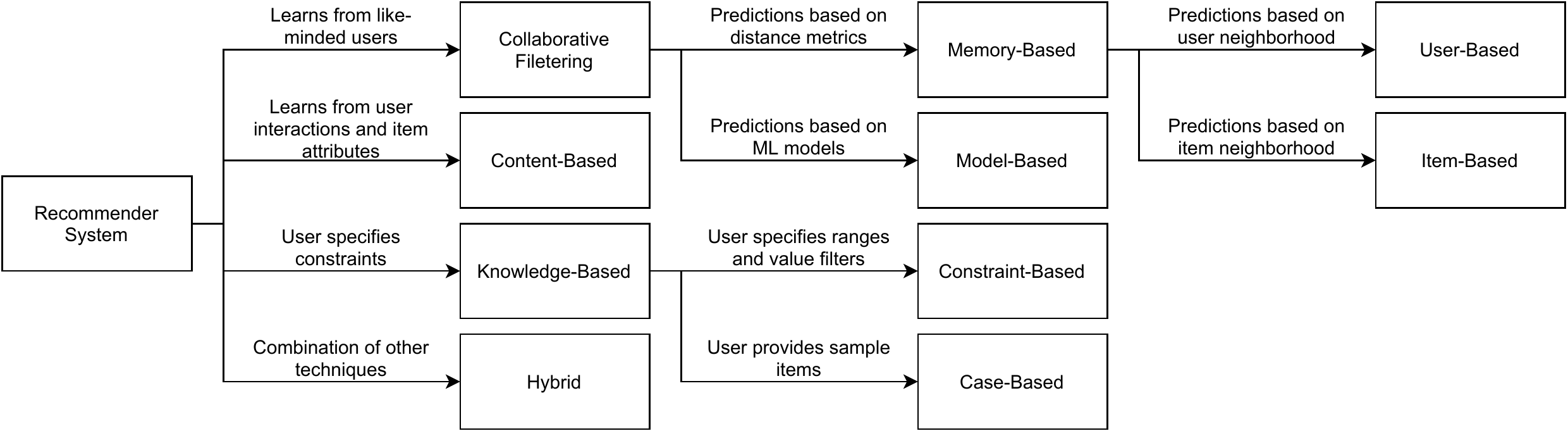}
	\caption{Taxonomy of Recommender Systems.}
	\label{fig:recsystaxonomy}
\end{figure}

A diagram with the most important types of Recommender Systems can be found in \cref{fig:recsystaxonomy}

\subsubsection{Recommender System Evaluation}

Before discussing evaluation metrics, it is important to make a distinction between how evaluations are made. \textit{Online Evaluation} methods present real users with different versions of recommender systems, and measure how each one performs. It is often the case that there are two versions and users are randomly split $50/50$, and this setting is called \textit{A/B Testing}. An automated generalization of this test setting can be implemented with \acrshort{rl}, using \textit{Multi-Armed Bandit} algorithms. Research using online methods is often limited, due to the commercial impact of testing unsuccessful recommenders, the lack of generalization of the domain of specific models, and commercial secrets. In contrast, \textit{Offline Evaluation} methods use historical datasets, often public, to train \acrshort{recsys} on recorded data. These are ideal for research because findings are comparable and reproducible.

There are several goals that can be used to compare Recommender Systems, and the following definitions are purposefully generic, since many researchers decide to calculate these concepts differently, based on the goals of their work. \textit{Accuracy} measures the percentage of the recommendations that would've been selected by a user. \textit{Coverage} measures the percentage of items that were recommended, out of all the items available. \textit{Confidence} measures the average statistical confidence or confidence interval (which some \acrshort{ml} models provide) in the recommendations. \textit{Novelty} measures the percentage of the current recommendations that has never been recommended to this user before. \textit{Serendipity} measures the percentage of the recommendations that would not have been selected by an obvious recommender (e.g., a simplistic model). \textit{Diversity} measures the similarity between the items that were recommended. \textit{Robustness} or \textit{Stability} are measurements of how a recommendation system can be affected by (i.e., how \textit{attack-resistant} it is to) false reviews or fake ratings. Lastly, \textit{Scalability} measures how an increased amount of items impact learning time, prediction time, and memory consumption.

An example of a concrete goal often used in research is to maximize \acrlong{ndcg} (\acrshort{ndcg}). This metric averages how relevant each item in the recommendations is to the user (Cumulative Gain). The averages are weighed by how early each item appears on the list (\textit{Discounted Cumulative Gain}), and normalized by how relevant each item is in general (i.e., the \textit{Ideal Relevance}). Given a user $u$, a slate of products $P = \{p_1, p_2, ..., p_N\}$, and a $relevance$ metric taken from user feedback, we can calculate the discounted cumulative gain as in \cref{eq:dcg}.

\begin{equation}
DCG_{u}(P) = \sum^{N}_{i=1} \frac{relevance_{u}(p_i)}{log_2(i + 1)}
\label{eq:dcg}
\end{equation}

Typically $relevance$ is the expected rating of a user to an item \cite[chap.~7.5.3]{recommendersystems}. Using ground-truth data (raw ratings) we can compute an ideal DCG metric, called $IDCG$, representing the maximum possible $DCG$. And finally, we have \cref{eq:ndcg}:

\begin{equation}
NDCG_{u}(P) = \frac{DCG_{u}(P)}{IDCG_{u}(P)}
\label{eq:ndcg}
\end{equation}

Which outputs a value in range $(0,1)$. Normally these metrics are reported as NDCG@N, where N is the number of products in the recommendation slate, or $|P|$.

\subsection{Deep Learning}

\acrlong{ml} is a field of Artificial Intelligence (like Natural Language Processing, or Computer Vision) that aims at building solutions to problems that involve pattern recognition, clustering, regression, classification, and decision-making. Regression and classification problems are commonly addressed by specific types of learning algorithms, and are referred to as \gls{supervisedlearning}. Neural Networks are a class of the latter, and \acrlong{dl} (\acrshort{dl}) refers to some specific types of those networks (i.e., when there are multiple hidden layers).

In Artificial Neural Networks (ANN), the main logic units are called \textit{neurons}, and are characterized by an activation function that combines a set of weights (and often biases) and produces an output signal \cite{deeplearning}. A \textit{perceptron} is a neuron in which the output is $1$ if the dot product of the weights is greater than $0$, and $0$ otherwise. This is often called a $linear$ function, and can be represented as in \cref{eq:linearfunction}:

\begin{equation}
  f(x) =
    \begin{cases}
      1 & \text{if $w \cdot x + b > 0$}\\
      0 & \text{otherwise}
    \end{cases}
\label{eq:linearfunction}
\end{equation}

where $w$ is the weight vector, $x$ is the input, and $b$ is the bias vector. Other examples of activations include \acrshort{relu}, the \textit{sigmoid} function, and the \textit{softmax} function.

The training of such learning models requires adjusting the weights in order to minimize a specific \textit{cost} function, which often is an aggregation of the \textit{loss} that occurred during training. Examples of cost functions include the Mean Square Error (MSE) or the \textit{Cross-Entropy} function. The logic that tells us how much to adjust the weights is called an \textit{optimizer}, and is often based on gradient descent optimization. The degree to which the optimizer moves in the direction of the gradient is referred to as \textit{learning rate}. Examples of optimizers include \textit{Momentum} and \textit{Adam}. The algorithm that optimizes all the neurons in a network accordingly is called \textit{backpropagation}.

When several layers of neurons are stacked together, the models are called \acrlong{dl}. The layers can be \textit{fully connected} or not, and \textit{architecture} refers to how connections are setup, how layers are created, the activation functions chosen for each neuron, and so on. There are several utilities for these methods, and they are discussed in more detail in \cref{sec:deeplearningrecsys}.

\subsection{Reinforcement Learning}

\acrshort{rl} refers to a class of \acrshort{ml} algorithms in which an \textit{agent} interacts with an \textit{environment}. The former tells the latter which action should be taken next, and receives a reward signal along with the next state. Commonly these interactions are measured by steps, and after some conditions are met the interaction episode ends (\textit{finite-horizon} environments). There are RL problems though, in which episodes go on continuously. In such settings it is customary to discount rewards, so that agents can learn actions that are good in the long-term, but so that rewards in the distant future tend to zero.

A classic example of a RL problem is the task of balancing a pole on a cart in two dimensions \cite{reinforcementlearning}. In this case, the environment would be a physics model, that would estimate the position, angle, and velocity of the pole. The agent would apply force to the cart, either left or right. For every step that the pendulum remains upright, the agent receives a $+1$ reward. If the pole gets too inclined, or the cart moves too far away from the starting position, the episode ends. By randomly exploring different strategies, many RL algorithms are able to learn how to balance the pole appropriately.

There is a notable class of algorithms, that seek to make optimal decisions without considering long-term agent-environment interactions (\textit{non-associative} tasks), called \textit{Multi-Armed Bandits}. Some examples of these include k-Armed Bandits, \acrlong{ucb} (\acrshort{ucb}) action selection, and Gradient Bandits. Bandit algorithms with environment context are called Contextual Bandits. 

Generally these algorithms start by randomly picking actions and observing the reward from each action. They estimate expectations of the reward, and greedily start choosing actions that maximize these expectations (this is called $\epsilon$-greedy exploration). \acrshort{ucb} methods have the advantage of modeling uncertainty in the reward expectation. Gradient Bandits go beyond and model the rewards as Boltzmann (\textit{soft-max}) distributions.

The more advanced \acrshort{rl} algorithms rely on the problems having certain attributes, specifically that they are \acrlong{mdp}es (\acrshort{mdp}). They are extensions of Markov Chains, and therefore require the Markov Property (\textit{memorylessness}). In other words, if we are at a particular state, the probabilities of transitioning to other states depend solely on information of the current state, not the past ones (i.e., a game of chess). Environments that don't provide agents with the full state are a special type of \acrshort{mdp}, called  Partially Observable Markov Decision Processes (POMDP).

Agents may learn policies, value functions, and models. Policies are a mapping between states and actions, that tells the agent which is the best action to take in each possible state. Policies can be deterministic or stochastic. Value functions predict the sum of all expected future rewards, if a particular policy is followed in the future (known as \textit{Return}). Models can be estimations of the transition probability between all the states, or estimated reward for all state-action pairs. RL agents that don't create models are referred to as \textit{model-free}. Some agents attempt to learn all value functions, some attempt to learn an optimal policy (sometimes known as \textit{Control}), and some attempt to do both. These last ones are called \textit{Actor-Critic} methods.

RL algorithms face several challenges that supervised models are not equipped to handle. The first one is that data is not \acrlong{iid} (\acrshort{iid}). The second problem is the trade-off between exploring new action or state scenarios, and exploiting previously obtained knowledge. In some cases actions the agent takes will only payoff later on, in other words, the expected reward can be delayed. Some problems also display the property of \textit{nonstationarity}, which means that the transition matrix or the reward distribution may change over time.

We can represent mathematically the value of being in a particular given state, using the \textit{Bellman equation}. It calculates the immediate reward, plus the discounted reward of all future interactions. Given a state $s$, the value $v_\pi(s)$ of being in this state if following a policy $\pi$ is defined as in \cref{eq:bellman}:

\begin{equation}
v_\pi(s) \doteq \sum_{a}\pi(a|s)\sum_{s',r}p(s',r|s,a)[ \,r + \gamma v_\pi(s')] \,
\label{eq:bellman}
\end{equation}

Here $\pi(a|s)$ is the probability of this policy choosing an action $a$ given state $s$, while $p(s',r|s,a)$ is the probability of receiving reward $r$ and transitioning to state $s'$ when taking action $a$ in state $s$. $\gamma$ is the discount factor for future rewards, and $v_\pi(s')$ is the recursive representation of the value of the following state $s'$ (often called a trajectory). This equation is solvable numerically, albeit often unfeasible due to the high dimensionality of the state space. RL algorithms are more efficient solutions to this problem.

Traditional RL algorithms seek optimal policies by storing a map of the value of every known state-action pair. These solutions are called \textit{Tabular Solution} methods. These are subdivided into \acrlong{dp} (\acrshort{dp}), Monte Carlo methods, and \acrlong{td} (\acrshort{td}) learning.

\acrshort{dp} methods are applied on environments that can be perfectly modeled, and they iteratively improve policies until reaching a mathematically optimal one. Examples of this collection of algorithms include \textit{Policy Iteration} and \textit{Value Iteration}. Monte Carlo methods don't require environment modeling, and are able to learn optimal policies from experience, by simulating different scenarios. An example of such a strategy is the \textit{On-policy MC control} algorithm. \acrshort{td} learning is a combination of the two previous ones, in the sense that it also learns from experience like Monte Carlo methods, and that experiences affect other related estimates, like in \acrshort{dp}. Examples of these algorithms include \textit{SARSA} and \textit{Q-Learning}.

\subsection{Deep Reinforcement Learning}

Tabular methods are very limited, when applied to real world problems. The first limitation is the need to model the state using discrete variables. A tabular solution would consider velocities of $14.999$ and $15$ to be two completely unrelated states. Another limitation is the unfeasibility of storing large numbers of states. A chess board for instance, has $10^{120}$ possible configurations, which is much more than we can currently store.

These problems can be solved with \textit{Approximate} solutions, that attempt to approximate towards an optimal policy, although convergence is not guaranteed in most cases. These solutions use supervised learning algorithms, and although other models could be used, there are several benefits of using \acrshort{dl}. There's the natural correlation between the data from RL environments, the changes in the value functions that happen when the policy changes, and the nonstationarity of many problems.

A very straightforward example of such models is a neuron with linear activation, in which inputs are the state features, and the output models the value of that particular state. The cost function is the mean squared error between the observed and predicted values. This algorithm is called \textit{Monte Carlo Gradient Descent}.

\subsubsection{Deep Q-Learning}

If we model the value of choosing each action $a$ in a particular state $s$, known as $Q(s,a)$, as an output neuron this algorithm is called \acrshort{dqn}. However, adding more neurons and layers introduces some challenges like the lack of \acrshort{iid} data, the value function changing after every update of weights, and the lack of exploration strategies. Several techniques have been developed by the \textit{Deep Mind} team at \textit{Google} to make \acrshort{drl} useful under \acrshort{mdp} problems. We will now discuss a number of the most important ones.

To simulate the \acrshort{iid} property researchers have developed the \acrlong{erb} (\acrshort{erb}), which stores a large amount of episode experiences prior to learning, and the agent samples randomly from it in small batches. We can use the loss (\acrshort{td}-error) from certain experiences to learn which among them moves the network the most along the gradient. We can use this information, known as \textit{priority}, to assign these experiences a higher probability of being sampled. When combining it with \textit{Importance Sampling} (to avoid bias towards high priority experiences) we have a \textit{Prioritized Experience Replay}, developed in \citeyear{schaul_prioritized_2016} by \citet{schaul_prioritized_2016}.

To handle the problem of the Q-value functions changing after every training, the concept of a \textit{target} network was introduced, which is a duplicate of the original network, but that gets updated every $k$ iterations. This means decisions are made using the \textit{target} network, but the original one gets updated during learning. If we obtain the Q-value from the \textit{target} network when learning we have \acrlong{ddqn} (\acrshort{ddqn}), developed in \citeyear{hasselt_deep_2016} by \citet{hasselt_deep_2016}.

\begin{figure}[!htbp]
	\centering
	\includegraphics[width=0.5\textwidth]{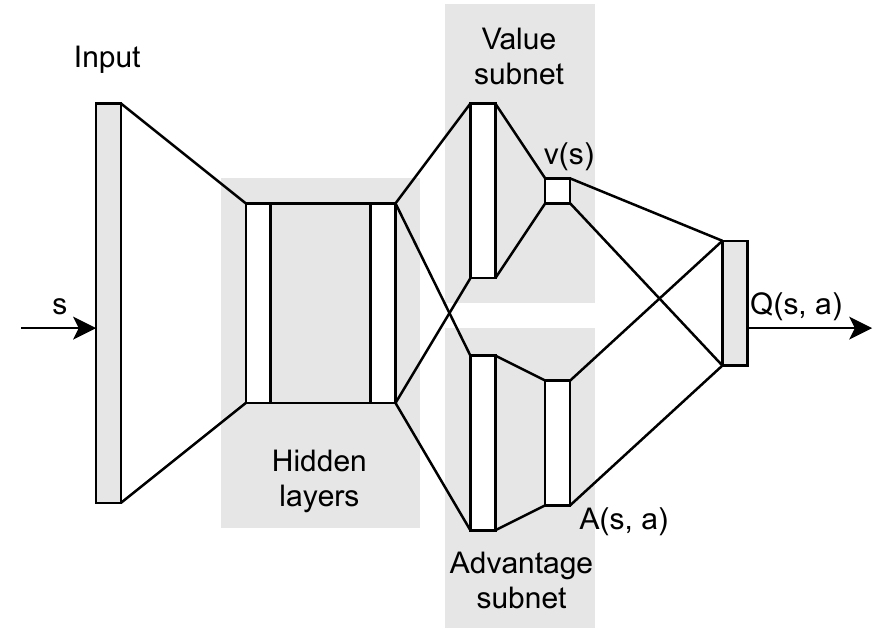}
	\caption{Example architecture of a Dueling Q-Network. Notice how value and advantage are modeled separately. Only one neuron is used to estimate $v(s)$, and to estimate $A$ and $Q$ we need as many neurons as there are actions available. The last layer is not a regular linear layer, but rather a calculation, as explained in \cref{eq:dueling}.}
	\label{fig:dueling}
\end{figure}

To handle the exploration problem, the $\epsilon$-greedy approach with a decaying exploration probability is often used. A more recent approach, called \textit{Noisy Nets}, is to introduce random noise in the hidden layers, developed in \citeyear{fortunato_noisy_2019} by \citet{fortunato_noisy_2019}. A recent approach to improve the estimations of the Q-value has been to split it into two different estimators: the value $v(s)$ of being in a state $s$ and the advantage of choosing an action $a$ in this state $A(a, s)$. This setup is called a \textit{Dueling Q-Network}, developed in \citeyear{wang_dueling_2016} by \citet{wang_dueling_2016} and an illustration of this concept can be found in \cref{fig:dueling}. The way $Q$ is calculated from $A$ and $v$ is by adding them, and subtracting the average advantage (to avoid the \textit{identifiability} problem).

\subsubsection{Policy Gradients \& Actor-Critic Methods}

One of the limitations of \acrshort{dqn} is that the action space needs to be necessarily discrete, since we estimate the value of each action using a neuron. A different approach is to model the output as a probability distribution of the expected return, estimated separately for each action. This obviously also works on discrete action spaces, by using \textit{softmax} distributions. Modeling this way means we will want to maximize the value outputs, instead of minimizing the cost. This \citeyear{silver_deterministic_2014} algorithm is a policy gradient version of \textit{REINFORCE} by \citet{silver_deterministic_2014}.

There are advantages and disadvantages of modeling this way. On one hand it is simpler, since there's no need for exploration, replay buffers, or target networks. On the other hand, the correlations between states of the same episode may introduce bias, the high variance of the reward signals can  cause slow convergence. 

Approximate Actor-Critic methods are a combination of the previous two approaches. The name comes from the fact that this method uses two distinct neural networks, an \textit{actor} that decides what action to take at each step, and a \textit{critic} that evaluates the action taken by estimating $Q$. \acrfull{a2c} are an example of this approach, but modeling the advantage of the state-action pair separately. They were developed in \citeyear{mnih_asynchronous_2016} by \citet{mnih_asynchronous_2016}.

\subsection{RL-Based Recommender Systems}

In the RecSys literature, there are two main ways in which RL is used. The first one is to create an automated way to decide which recommender system (out of a pool of options) should be used with which user. The second way is to use item features to deliver personalized recommendations to each user. In both cases, most of the literature focuses on tabular methods such as \textit{Contextual Bandits}, the $\epsilon$\textit{-greedy Algorithm}, or the \acrlong{ucb} algorithm (\acrshort{ucb}). Research in solving the \acrshort{recsys} problem using \acrshort{rl} is very limited, and there is a lot of opportunity in this field \cite[see][chap.~13.3]{recommendersystems}.

The RL literature refers to the second usage as \textit{Personalized Web Services} \cite[see][chap.~16.7]{reinforcementlearning}. In such setting agents use feedback directly (or via user clicks) to iteratively improve a recommendation policy. This is often referred to as \textit{Associative Reinforcement Learning}, and it has had success in improving clickout-rates of online sellers, and has been shown to have better results when modeled as a \acrshort{mdp}. One of the potentials of using RL in this scenario is the long term effect (also known as \acrlong{ltv} or \acrshort{ltv})  of a policy, having the potential to convince a user to purchase something through a sort of \textit{recommendation funnel}. Researchers testing RL models on commercial solutions often use \textit{off-policy} evaluation to avoid financial risks.

\section{Current Technologies}
\label{sec:currenttechnologies}

\subsection{Deep Learning And Recommender Systems}
\label{sec:deeplearningrecsys}

Since the application of \acrshort{rl} in the context of \acrlong{recsys} is fairly new, there are not many review papers available. Nevertheless, two important review articles are worth discussing.

The first article from \citeyear{deeplearningrecsysreview} reviews the most important research in using \acrshort{dl} to build \acrshort{recsys}, by \citet{deeplearningrecsysreview}, and briefly discusses some research that used \acrshort{drl}. It explains how several tech companies use \acrshort{dl} to enhance recommendation quality. Some examples of such systems are recommendations in the form of YouTube videos, Google Play apps, and Yahoo news articles. All of these having  served billions of user sessions over the years. Naturally, the amount of academic research around \acrshort{dl} and \acrshort{recsys} grows steadily.

There are several benefits of using DL. Firstly the ability to model problems nonlinearly, which is a property that enables these models to recognize more complex patterns. Secondly there's representational learning, which allows large number of features, as well as less traditional ones like images, audio, or text, to be used in the learning process. Finally there's the ability to model signals sequentially, which in some cases allow Recurrent Neural Networks (RNN) to represent some temporal dynamics. Despite these benefits, \acrshort{dl} is not without some setbacks. These models often make the interpretability of recommendations very difficult, require large amounts of data to be trained, and extensive hyperparameter tuning.

It is important to look at the latest advances in \acrshort{dl} because there are many important concepts of it that could be successfully applied in the field of \acrshort{drl}. The concept of \textit{autoencoders}, for instance, could be used for dimensionality reduction in finding latent variables in \acrshort{rl} state vectors. \textit{Attention mechanisms} could be used to improve recommendation interpretability. \textit{Convolutional layers} could be used to introduce images to environments.

In spite of all this potential, \acrshort{dl} by itself has no means of optimizing recommendations in the long term, nor it has the ability to perform well on nonstationary scenarios (user behavior changes over time), nor the ability to handle mutating product catalogues. All of these are often requirements in recommender systems, and they're where \acrshort{rl} becomes very relevant.

The second review article worth mentioning is an overview of \acrshort{rl} (specifically \acrshort{mdp}) techniques applied to \acrshort{recsys} \cite{gupta_study_2018}. These studies validate the usage of partially observable environments, \textit{off-policy} training, nonlinear functions, the $\epsilon$-greedy algorithm, \textit{SARSA}, \textit{value iteration}, and \acrlong{dqn}. The most important reviewed works are discussed in \cref{sec:rlrecsys}.

\subsection{Reinforcement Learning And Recommender Systems}
\label{sec:rlrecsys}

We'll review chronologically some the most cited journal articles and conference papers, in the field of applying \acrlong{rl} to the \acrshort{recsys} problem. Then, we look at some interesting \acrshort{rl} environments, and details about the most recent ACM \acrshort{recsys} conference.

We start by analyzing the \citeyear{mahmood_learning_2007} work of \citet{mahmood_learning_2007}. They showed how Conversational Recommender Systems could be modeled as \acrshort{mdp}. Their system was based in users searching travel products, and relied on them specifying constraints. In this case, the \acrshort{rl} agent suggests improvements to the constraints when too many or too little recommendations were returned. In their model, there is a very limited number of states an agent can be in, and a limited number of actions it can take. They show how an agent could improve recommendation policies by using the Policy Iteration algorithm, and evaluate it using simulations. In \citeyear{mahmood_improving_2009} \cite{mahmood_improving_2009} they validate their model in an online setting, by comparing it with a fixed recommendation policy. They found that users that received recommendations from the \acrshort{rl} agent reached their goal faster, more often, and adopted more query improvement suggestions.

Next we analyze the research from \textit{Yahoo! Labs} in recommending news articles. In \citeyear{li_contextual-bandit_2010} \citet{li_contextual-bandit_2010} modeled a news recommender system as a contextual bandit problem that maximizes user clicks on articles. They show that it's possible to reliably train \acrshort{rl} models offline with tracking data, recorded by recommending articles randomly. They deploy the trained model online, and show that it performs better than a multi-armed bandit agent without user context. In \citeyear{li_unbiased_2011} \citet{li_unbiased_2011} addressed some problems of training agents offline, specifically the modeling bias often present in user simulations. It was accomplished by introducing a replay methodology (analogous to an \acrlong{erb}), and research shows how training offline by replaying dataset experiences can produce comparable results to online training. Results were measured in terms of \acrshort{ctr}. 

Two more theoretical works from \textit{Yahoo! Labs} helped prove regret bounds for a few models. In \citeyear{kale_non-stochastic_2010} \citet{kale_non-stochastic_2010} showed how a multi-armed bandit model could be adapted to provide a set of actions per step. They analyze scenarios in which the actions need to be ordered, and scenarios that don't require ordering. They provide theoretical a bound for the regret metric of such agents. In \citeyear{chu_contextual_2011} \citet{chu_contextual_2011} provided a theoretical analysis of a variant of \textit{LinUCB}. They prove a high-probability bound for the regret of their agent. There is a discussion about how to deal with scenarios when these linear models are not guaranteed to accurately estimate the expected reward of recommendations. It may be the case that in real online systems often it's not possible to perfectly obtain the expected reward, especially in non-stationary settings.

This is exactly the problem that \citet{bouneffouf_contextual-bandit_2012} address in \citeyear{bouneffouf_contextual-bandit_2012}. They develop an $\epsilon$-greedy model in which the exploration probably $\epsilon$ depends on the user's situation. The precise definition of situation is the similarity between a user's feature vector, and the closest vector in its history of features. They evaluate their model offline, using a dataset from a commercial system, and measuring \acrshort{ctr}. They compare it to a standard $\epsilon$-greedy agent, and one with decreasing exploration probability, and show that their model performs best.

Now we review important work from \textit{Adobe Research} in the area of ad recommendation. In \citeyear{thomas_high-confidence_2015} \citet{thomas_high-confidence_2015} studied effective ways to evaluate recommendation policies offline. They discuss the motivation, and the high risks of using a poor recommendation strategy in live systems. Their evaluation technique accurately estimates a lower confidence bound on the expected return of a given policy. Their methodology is evaluated using real recorded \textit{Adobe Marketing Cloud} data, and training a user simulation from it. It's shown how the confidence interval for predictions narrows down when the volume of training data increases, as would be expected. Also in \citeyear{theocharous_ad_2015}, \citet{theocharous_ad_2015} investigate the benefits of training \acrshort{rl} agents to increase long-term user interactions (known as \acrlong{ltv} or \acrshort{ltv}), as opposed to maximizing \acrshort{ctr}. They also provide means of evaluating such models offline, to enable hyperparameter optimization, and to guarantee safe deploys of these policies. They use a \textit{Fitted-Q Iteration} RL agent, that in turn uses a combination of $\epsilon$-greedy exploration and a Random Forest classifier. Even though this model handles highly dimensional states, it does not solve the problem of a large action space.

The work that \citet{choi_reinforcement_2018} did in \citeyear{choi_reinforcement_2018} was really important in addressing highly dimensional data, both in the state space and in the action space. They introduce a biclustering step before training, that drastically reduces the amount of states and actions of RL agents. Two added benefits are that clustering actions allows some explainability of recommendations, and clustering users allows some generalization of recommendations, which effectively handles the \gls{coldstartproblem}. They test Q-Learning and SARSA offline, on the \textit{MovieLens} movie recommendation datasets. It would have been interesting to see comparisons of other dimensionality reduction techniques, and non-tabular agents.

Lastly, we look at recent works that used \acrshort{drl}. \citet{zheng_drn_2018} in \citeyear{zheng_drn_2018} used a dueling \acrshort{ddqn} to create a news recommendation system. With their model, they wanted to improve agents that optimized for short-term goals, that relied solely on \acrshort{ctr}, and that had low Diversity. They evaluate that their model achieves these goals both with offline training, and online. These works give us an idea on how a utility matrix can be adapted, and neural networks can be used to predict its metrics. By modeling the network input (the state) as a combination of features (user, item, interaction, context), the output can be, for instance, the probability of this user clicking on this item at some point in the future (in other words, the expected click-out rate).

\citet{chen_stabilizing_2018} in \citeyear{chen_stabilizing_2018} address the problem of high variance and biased expectation of rewards, typical of recommender systems. To improve the variance they introduce a modification to the \acrshort{erb}, by stratifying stable user features (like age or gender). This means sampling experiences will draw similar amounts of experiences from these strata. To improve the expectation bias, they propose the concept of \textit{Approximate Regretted Reward}, in which the reward of an agent learning online is penalized by the reward of an agent that learned with previously available offline data. They apply these concepts to a \acrshort{ddqn} model, and test it on a live search engine. Their model had over $20$ thousand possible actions (user suggestions) and the states were modeled after user features. It's shown to perform better (in terms of \acrshort{ctr} and conversion rate) than a system based on a \gls{supervisedlearning} algorithm, on an A/B test.

\citeauthor{zhao_recommendations_2018} published two important works in \citeyear{zhao_recommendations_2018}. The first one \cite{zhao_recommendations_2018} is about a model that explicitly requires users to vote if a recommendation was useful or not. The agent learns directly from these reward signals. They tested it with data from an online e-commerce platform. The second one \cite{zhao_deep_2018} addresses some problems of learning with real-time user feedback, and recommending pages of products. They develop a page-wise Actor-Critic \acrshort{rl} agent that is able to maximize clicks on a 2D grid of product recommendations. The researchers train and test the model offline, but also perform online evaluations.

\subsection{Simulation Environments}

With the recent advances in \acrlong{rl} and \acrshort{ml} in general, certain tools and frameworks start to become standards. One of the most influential \acrshort{rl} benchmark tools is the \citeyear{openaigym} \textit{OpenAI Gym} by \citet{openaigym}. We will therefore explore research that was done in designing gym environments that specifically simulate the dynamics of recommendation systems.

We start by reviewing \textit{Google Research}'s \citeyear{ie_recsim_2019} RecSim by \citet{ie_recsim_2019}. It consists of a gym environment in which it's possible to flexibly define models for users, documents (items), and user-choice interactions. User models refer to the features of a user (e.g., demographic data, context), and its probability distributions. The same modeling (sampling features from a probability distribution) is done for documents. The probability of a user selecting a recommendation is also sampled from a distribution, and so are the transitions that may occur in a user's state when it selects a document (e.g., a user's engagement may increase or decrease). The agent's task is to optimally recommend documents to users. RecSim comes with several already prepared environments.

A different approach called PyRecGym was researched in \citeyear{shi_pyrecgym_2019} by \citet{shi_pyrecgym_2019}. Also a set of gym environments, this framework uses some traditional \acrshort{recsys} datasets (e.g., MovieLens, Yelp) to create simulations of recommender systems. The tool divides datasets into \textit{initialization data} used to train user simulators, and \textit{interaction data} used to teach the RL agent. User and item features are combined, and recommendation selections by the simulated user happen only if they occurred in the initialization dataset. This is a naive way of modeling human-computer interactions, and this environment would have benefited from a probability distribution when simulating the interactions.

Another gym environment that is worth reviewing is the \citeyear{santana_mars-gym_2020} MARSGym by \citet{santana_mars-gym_2020}. This environment is designed specifically for simulating marketplace dynamics, in which recommendation systems not only need to optimize click-throughs, but also maintain healthy levels of fairness (recommendation diversity). It bears similarities with PyRecGym in the sense that it learns from tracking data, and that user-interactions simulations are exact replicas of what happened in the historical data. The researchers go into detail about the problems and benefits of offline training, and how counterfactual estimators can be used to reduce the bias. They demonstrate the effectiveness of the environment using data from \textit{Trivago}.

\subsection{The ACM RecSys Conference}

To conclude this review, it's important to investigate not only in academic and commercial production, but also in competitions. In \citeyear{jannach_why_2020} \citet{jannach_why_2020} did an extensive exploration of algorithms used in competitions, and share their findings and hypothesis. From the top 5 winning teams of the ACM RecSys Conference from 2017 to 2019, only one used \acrshort{dl}. All the others used extensive feature engineering and more traditional \gls{supervisedlearning} algorithms. A similar situation happens in other related competitions. The authors argue how competition datasets are more massive than the ones used in academic research, and how the time available to train models in competitions is often short. \acrshort{dl}-based models often require more training time, more expensive hardware, and more extensive hyperparameter optimization. Another important aspect is that competition datasets tend to be much more sparse, which causes overfitting in many \acrshort{dl} models.

Some of the authors of this paper were part of the winning team of the 2020 RecSys challenge, and their methodology is described in \citet{schifferer_gpu_2020}. The greatest contributions of this paper are their efforts in improving the performance of preprocessing, feature selection, and model training on huge datasets. They are able to run all these steps in the GPU, and speed up computations $280$ times, reducing processing time from several hours to a couple minutes. The authors describe how they compared several deep learning models and other supervised ones, but \textit{XGBoost} was the one that performed best.

\chapter{Materials And Methods}
\label{chapter:procedure}

\section{Learning Environment Selection}

In the previous chapter we proposed to evaluate three modern \textit{gym} environments found in the literature: PyRecGym \cite{shi_pyrecgym_2019}, RecSim (the interest evolution environment) \cite{ie_recsim_2019}, and MARS-Gym (the Trivago Rio environment) \cite{mars}. Unfortunately, it proved impossible to find an open-source implementation of PyRecGym, therefore we will replace it in this comparison with one of Google's ML Fairness Gym \cite{fairness_gym} recommender environments. ML Fairness Gym is based off RecSim, and we'll be discussing specifically the experimental source code based on the 1M MovieLens dataset \cite{harper_movielens_2015}.

When comparing these environments there are many desirable traits we are looking for. The usage of real-world data, for instance, makes the environment more relevant to realistic problems. The availability of large amounts of data makes it possible to properly train deep neural networks. The documentation of the environment is also important, to enable us to understand the features of the states. The time it takes for a time-step to be simulated is also paramount, because smaller execution times allow for further experimentation. In \cref{table:rl_environments} these traits are displayed in comparison.

\begin{table}
\centering
\begin{tabular}{l}
\begin{filecontents*}{csv/rl_environments.csv}
,MARS Gym,RecSim, ML Fairness Gym
Simulatable episodes,1,Infinite,6040
Real-world data,Yes,No,Yes
Documentation,Lacking,Comprehensive,Lacking
Time-step avg. duration,0.14 ms,1.16 ms,0.33 ms
Models user interest,No,Yes,Yes
User features,None,Numeric (20),None
Item features,Numeric (148),Numeric (20),Binary (19)
Interaction Features,Binary (779), Miscellaneous (5), Categorical (1)
\end{filecontents*}
\csvautotabular{csv/rl_environments.csv}
\end{tabular}
\caption{Comparison of different Recommender System environments.}
\label{table:rl_environments}
\end{table}

MARS-gym only has one possible episode because its definition of an episode is a sequential iteration over its ground-truth dataset (exactly like a supervised learning training routine). Meanwhile on the others, episodes represent a user's journey. The state of an environment's documentation specifically relates to the existence of explanations regarding the RL variables (i.e., state/observations, actions, reward, and the done flag). We benchmarked the time taken to simulate a time-step, averaged over 50 episodes, and taken using the same hardware (2.4GHz x 16 Intel i9). Different machines will of course output different simulation times.

Since none of the studied environments are suitable for this investigation, our decision will be to use the ML Fairness environment, but extending it to incorporate MovieLens user features such as sex, age, occupation, and zip code. Another benefit of this environment is the extensive usage of the MovieLens dataset in \acrshort{recsys} literature, which allows for comparison with other works. There's also the possibility of implementing environments using other recommendation datasets (i.e., a Trivago-based environment).

\section{How RecSim Simulates Recommender Systems}
\label{section:recsimsimulation}

As previously discussed, the RecSim framework (which our environment is based on) contains $4$ key abstractions of recommender systems, that need to be configured:

\begin{itemize}
  \itemsep0em 
  \item A \textit{User Model} that defines ways in which users are generated. This logic specifies which features are part of the user, and how they are generated (sampled from a distribution, or read from a dataset). In our case, this includes user features from the MovieLens data (i.e., sex, age, occupation, zip code).
  \item A \textit{Document Model} that defines ways in which documents are generated. Like the user model, this logic specifies which features describe an item. In our case it includes features from the MovieLens dataset, merged with violence scores from the Genome project \cite{vig_tag_2012} (i.e., title, genres, year, violence score).
  \item A \textit{User-Choice Model} that simulates a user's response to a particular document (i.e., a five-star rating to a movie). Like the user model, this logic specifies which features describe an item. This can be generated for example by sampling from a distribution, or by reading a matrix factorization of a user-interaction dataset. User-Choice models may also contain features like timestamps, user interactions, and so on. In our case this data comes from a matrix factorization of the MovieLens ratings.
  \item A \textit{User Transition Model} that determines how user features are affected after each user-document interaction. For instance, a user's interest in a certain retail category may decrease after interacting with it. In our case, we've configured users to slowly get addicted to certain movie categories; this is simulated by slowly increasing user affinities (their embeddings obtained by the matrix factorization).
\end{itemize}. 

Additionally, it allows the configuration of recommendation \textit{slates}. A slate is a subset of all the available documents that is presented to the user, and the \acrshort{rl} agent is responsible for deciding which documents are included in the slate. This type of strategy is often seen in e-commerce websites under sections that display items similar to the one currently being seen by the user (e.g., ``You may also like''). During the execution of the episodes, documents and users can be sampled randomly, with or without resampling, or divided into pools (i.e., users may be divided into \textit{train}, \textit{test}, and \textit{validation} pools).

Scores are simulated using a matrix decomposition strategy, from the utility matrix. Given the utility-matrix $U$ of shape $M\times N$, where element $U_{i,j}$ represents the rating, in range $(0,5)$, provided by a user $M_i$ to a movie $N_j$. Since $U$ is often sparse, and we need to simulate any user-movie recommendation, we can use a non-negative Single-Value Decomposition (SVD) algorithm that replaces missing ratings with the average before training. This is a traditional recommendation strategy that has gained a lot of attention since the 2006 Netflix challenge \cite{koren_matrix_2009}. The technique consists of estimating matrices $(W, H)$ such that $W * H \approx U$. We can then estimate a user's rating of an item by the dot product of a row of W (user embeddings) and a column of H (document embeddings). The larger the number of components the more accurate the factorization, and since the ML-Fairness project uses $55$ components, we will use the same amount. 

In \acrshort{rl} terms, the action space of this environment comprises of the 3883 movies that can be recommended at any moment. The observation space, just like in RecSim environments, is composed of the document's (movies) features, the user's response to the last seen item, and the user's features. The movies in this environment have 19 binary features, that represent the movie's category (e.g., drama, adventure). The features of the response are a simulated 5-star rating, and a violence score (extracted from the Genome dataset). The reward signal is in the range $(0,1)$ and relates to the user's simulated rating of the recommended movie. The reward is a multi-objective model, and represent a trade-off between interest and violence. A user's journey (history) $H_t$ at time-step $t$ can be represented as: $$H_t = [O_0, A_0, O_1, R_1, A_1, ..., A_{t-1}, O_t, R_t]$$where $O_i$ contains the features of available documents, current features of the user, and features of the latest user-document interaction at time-step $i$. $A_i$ represents the recommendation of a slate of movies at time-step $i$, and $R_i$ represents the reward received by the agent at time-step $i$. The embeddings used to generate the reward are not observable by the agent, as this is partly what the \acrshort{rl} agent tries to estimate.

\section{Comparison Metrics}

As described in \cref{section:recsimsimulation} the \acrshort{rl} models will be maximizing the expected return (i.e., the simulated user ratings). Several other studies measure different metrics, that may for instance help assess the health of a recommendation ecosystem, such as diversity. Modeling a return that balances these goals, performing a counterfactual evaluation, is a field under active research. Martin Mladenov (personal communication, May 5 2021), one of the authors of RecSim, explained there are hard methodological challenges involved in counterfactually accurate simulations.

Other studies measure the efficiency of a \acrshort{recsys} using traditional metrics like NDCG from \cref{eq:ndcg}, but depending on how relevance is modeled, there are no guarantees that the highest expected reward will output a high NDCG. Some researchers, instead of the expected reward, simplify the simulations using clickthrough-rates or hit-rates. Arguably, simulating reviews provides more accurate simulations, given the higher cardinality of simulated user evaluation. Below is a brief description of recent studies that use \acrshort{rl} and the MovieLens dataset, and \cref{table:mlstudies} summarizes them.

\citet{afkhamizadeh_automated_nodate}, measured their agent using cumulative custom regret metric, but it proved hard to understand how it's calculated and draw comparisons. Their best model achieved a cumulative regret of approximately $30$ after $100$ time-steps.  

The \citeyear{huang_deep_2021} study by \citet{huang_deep_2021}, measured not the expected return, but percentage of recommendation slates that contained a movie the user saw (i.e., Hit-Rate, or HR@N where $N$ is the slate size). Their best results are approximately $45\%$ HR@10 for the MovieLens1M dataset, after training for an unknown number of steps. \citet{sanz-cruzado_simple_2019} in \citeyear{sanz-cruzado_simple_2019} showed on the same dataset a cumulative recall of approximately 0.6 after 3 million time-steps, for $\epsilon$-greedy, k-NN with $k=10$, and kNN-based bandits. Another \citeyear{zou_reinforcement_2019} study by \citet{zou_reinforcement_2019} found NDCG@3 (from equation \ref{eq:ndcg}) of $75\%$ using REINFORCE on the same dataset after training on $90\%$ of the dataset (around 900k time-steps). 

Another study from \citeyear{yu_interactive_2020} that uses the same style of measurement (but they call it \acrshort{ctr}) by \citet{yu_interactive_2020} reported $35\%$ in $30000$ time-steps, using a Trust Region Policy Optimization (TRPO) algorithm. They experimented with Q-Learning but found TRPO to wield better results.

The \citeyear{liu_top-aware_2020} study by \citet{liu_top-aware_2020} also used a similar metric, although they called it Precision, and tested several models. One of the best performing models was DQN, and they found $65\%$ Precision@20 after training on $80\%$ of the user interactions (around 800k time-steps). Very similar previous works by the same authors \cite{liu_state_2020} \cite{liu_end--end_2020}, on the same dataset, and using the same DQN model, had found $54\%$ Precision@20.

\begin{table}[!htbp]
\centering
\begin{tabular}{*6l}\toprule
    Year & Metric & Best Result & Timesteps & Reference \\\midrule
    \begin{filecontents*}{csv/mlstudies.csv}
year,metric,best,timesteps,reference
2017,Cum. Regret,30,100,\cite{afkhamizadeh_automated_nodate}
2019,Cum. Recall,0.6,3M,\cite{sanz-cruzado_simple_2019}
2019,NDCG@3,0.75,900K,\cite{zou_reinforcement_2019}
2020,CTR,0.35,30K,\cite{yu_interactive_2020}
2020,Precision@20,0.8,800K,\cite{liu_top-aware_2020}
2020,Precision@20,0.54,800K,\cite{liu_state_2020}
2021,HR@10,0.45,N/A,\cite{huang_deep_2021}
\end{filecontents*}
\csvreader[head to column names, late after line=\\]{csv/mlstudies.csv}{}%
       {\year& \metric& \best& \timesteps& \reference& }%
    \bottomrule
\end{tabular}
\caption{Comparison of recent studies that train RL agents using the MovieLens dataset. Note the diversity of metrics.}
\label{table:mlstudies}
\end{table}

\section{Evaluation Agents}

The evaluation of the performance of our \acrshort{rl} agents will be done using three distinct agents: a Dueling DQN implementation that uses \acrshort{per} buffers and noisy networks, a REINFORCE implementation using discounted episode rewards as a baseline, and an Actor-Critic implementation that uses a value estimator as the critic. The source code will be written in \textit{Python}, with tools such as \textit{Jupyter notebooks}, \textit{pandas}, and \textit{Pytorch}. The code is made publicly available on the source-control management platform \textit{Github}\footnote{\href{https://github.com/luksfarris/pydeeprecsys}{github.com/luksfarris/pydeeprecsys}} and the notebooks are provided for reproducibility.

The Dueling DQN neural network is fully connected with one input neuron for each observation variable (there are 25), hidden noisy layers with linear neurons and \acrshort{relu} activation, a noisy value-estimator subnetwork with \acrshort{relu} activation, a noisy advantage-estimator subnetwork with \acrshort{relu} activation, and the noisy output layer with one neuron for each movie (currently there are 3883). The hidden layers are model parameters. The network uses the \textit{Adam} optimizer, and the learning rate $\eta$ is a parameter. The noisy layers use gaussian noise that helps with exploration, and the starting noise $\sigma$ is a parameter. The discount factor $\gamma$ used to calculate the \acrshort{td}-error is also a parameter. The \acrshort{per} is also parametrized with the buffer size, the burn-in, the batch size, the priority importance $\alpha$, the weight effect $\beta$, the $\beta$ annealing coefficient, and the minimum allowable priority $\epsilon$. The frequencies used to train the main network, and to update the target network, are also parameters of the model. The pseudo-code is demonstrated in \cref{algo:dueling}.

\begin{algorithm}[!ht]
\SetAlgoLined
 Let $Q_{\theta}$ be the main network with weights $\theta$ and noisy layers with noise $\sigma$\;
 Let $Q'_{\theta'}$ be the target network\;
 Let PER be the prioritized experience buffer of size $N_D$, burn-in $N_{in}$, batch size $N_{batch}$, priority importance $\alpha$, weight effect $\beta$, annealing $\beta_a$, and minimum priority $\epsilon$\;
 Let $P$ be the priorities of the experiences of PER\;
 Let $F_{train}$ be the frequency at which the main network is trained\;
 Let $F_{sync}$ be the frequency at which the target network is updated\;
 Let $c$ be the step count\;
 Let $\eta$ be the learning rate\;
 \For{each Time-Step $t$, $S_t$, $\gamma_t \in (0,1)$}{
 	Let $ready \leftarrow (|PER| > N_{in})$\;
 	Let $A_t$ be $argmax\,Q'_{\theta'}(S_t)$ if $ready$ else a random valid action\;
	Let $R_t$, $S_{t+1}$ be the environment's reward and next state for $A_t$\;
 	\If{$ready$}{
		$c \leftarrow c + 1$\;
		\If{$c \bmod F_{train} = 0$}{
			Let $\Pi \leftarrow P^{\alpha}$\;
			Let $b$ be a randomly sampled batch of size $N_{batch}$ from $PER$ using probabilities $\Pi$\;
			\tcp{Importance Sampling}
			Let $w^{b} \leftarrow (\frac{1}{N_{batch}} \frac{1}{P_b})^\beta$ be their weights\;
			\tcp{Weight annealing}
			$\beta \leftarrow min(\beta + \beta_a, 1)$\; 
			\tcp{TD Error following the DDQN update}
			Let $\delta \leftarrow Q_{\theta}(S^b, A^b) - \gamma_t\,Q'_{\theta'}(S'^b, argmax\,Q_\theta(S^b)) + R^b$\;
			\tcp{Weighed mean square errors}
			Apply gradient descent $\theta \leftarrow \theta\,\eta\,\frac{(\delta^2\,w^b)}{|b|}\,\nabla_{\theta}\,Q_{\theta}(S^b, A^b)$\;
			$P_i \leftarrow \delta_i + \epsilon$ for every experience $i$ in $b$\;
		}
 		\If{$c \bmod F_{sync} = 0$}{
			\tcp{Target network update, copy the weights $\theta$ over to $Q'$}
			$\theta' \leftarrow \theta$\;
		}
	}
	Store $(S_t$, $S_{t+1}, A_t, R_t)$ in PER with priority $max(P)$\;
 }
 \caption{Dueling DQN with Prioritized Experience Buffer and Noisy Layers}
 \label{algo:dueling}
\end{algorithm}

The REINFORCE model does not use noisy layers, uses \textit{tanh} activation, and \textit{softmax} output neurons. The hidden layers and learning rate $\eta$ are parameters of this model. The discount factor $\gamma$ used to calculate the baseline is also a parameter. The network is trained at the end of each episode. Predictions are made sampling from the predicted multinomial probability distribution, meaning the policy is stochastic.

The Actor-Critic agent includes the same parameters from the REINFORCE agent, namely the learning rate $\eta_{actor}$, the actor hidden layers, and the discount factor $\gamma$. Additionally, it includes the critic's hidden layers and learning rate $\eta_{critic}$. The critic's network is fully connected and it uses \acrshort{relu} activation. Both networks are trained at the end of each episode.

Regarding the loss calculations and subsequent backpropagation, the Dueling DQN agent uses the mean of the squared \acrshort{td}-errors of the batch, multiplied by their \acrshort{per} weights. For more detailed representations of the network architectures and gradient calculations, please refer to the Appendix \ref{ap:architectures}. As in DDQN agents, the \acrshort{td}-error is calculated using the target network to estimate the expected $Q$-value of the next state. The feedforward logic that combines the estimated value and advantage is detailed in \cref{eq:dueling}.

\begin{equation}
Q(s,a) = V(s) + Adv(s,a) - \frac{1}{|\mathcal{A}|} \sum_{a' \in \mathcal{A}} Adv(s,a')
\label{eq:dueling}
\end{equation}

The subtraction of the average advantage is needed due to the identifiability problem (i.e., decomposing $Q$ into $V$ and $Adv$).

While the critic error is the \acrshort{td}-error, the REINFORCE and Actor loss functions are the application of the Policy Gradient Theorem, as expected, and are described in \cref{eq:policygradient}.

\begin{equation}
\nabla_{\theta}J(\theta) = \mathbb{E}_{\pi}[G_t \nabla_{\theta} ln \pi_{\theta}(a|s)]
\label{eq:policygradient}
\end{equation}

\section{Evaluation Methods}

Due to the large number of parameters in each agent, performing an exhaustive hyperparameter space search would not be efficient. To that effect, following the methodology from \cite{reinforcementlearning}, we will test one parameter at a time, freezing the remaining ones. Each combination is tested $3$ times for $500$ episodes, and the average of the final moving mean (100 episode window) of the Return will be used for comparison. The best value found for each parameter is chosen in the final model.

Once models have the best parameters found, we run the comparison for 5 different seeds and $2000$ episodes. The moving mean will be used to compare the agents, along with its $95\%$ confidence interval. A random recommendation baseline is included for comparison. After the best agent is chosen, we run a training with slates of size $10$ to compute the \acrshort{ndcg} metric. This runs again for 5 seeds, and the agent is trained for $500$ episodes (25K time-steps).

\section{Implementation Details}

The source-code used to run the tests was developed using the \textit{git} source control technology, and stored online and made available publicly on the \textit{Github} website. The needed changes made to the MLFairness repository were made in a fork, available as a \textit{submodule}. The repository is licensed under the MIT license, which allows the code to be used privately and commercially, but without liabilities or warranties. The repository contains \textit{wiki} pages with detailed instructions on installation and execution. Additionally, the repository contains \textit{jupyter notebooks} with example usage of the environment.

\begin{figure}[!ht]
	\centering
	\includegraphics[width=1\textwidth]{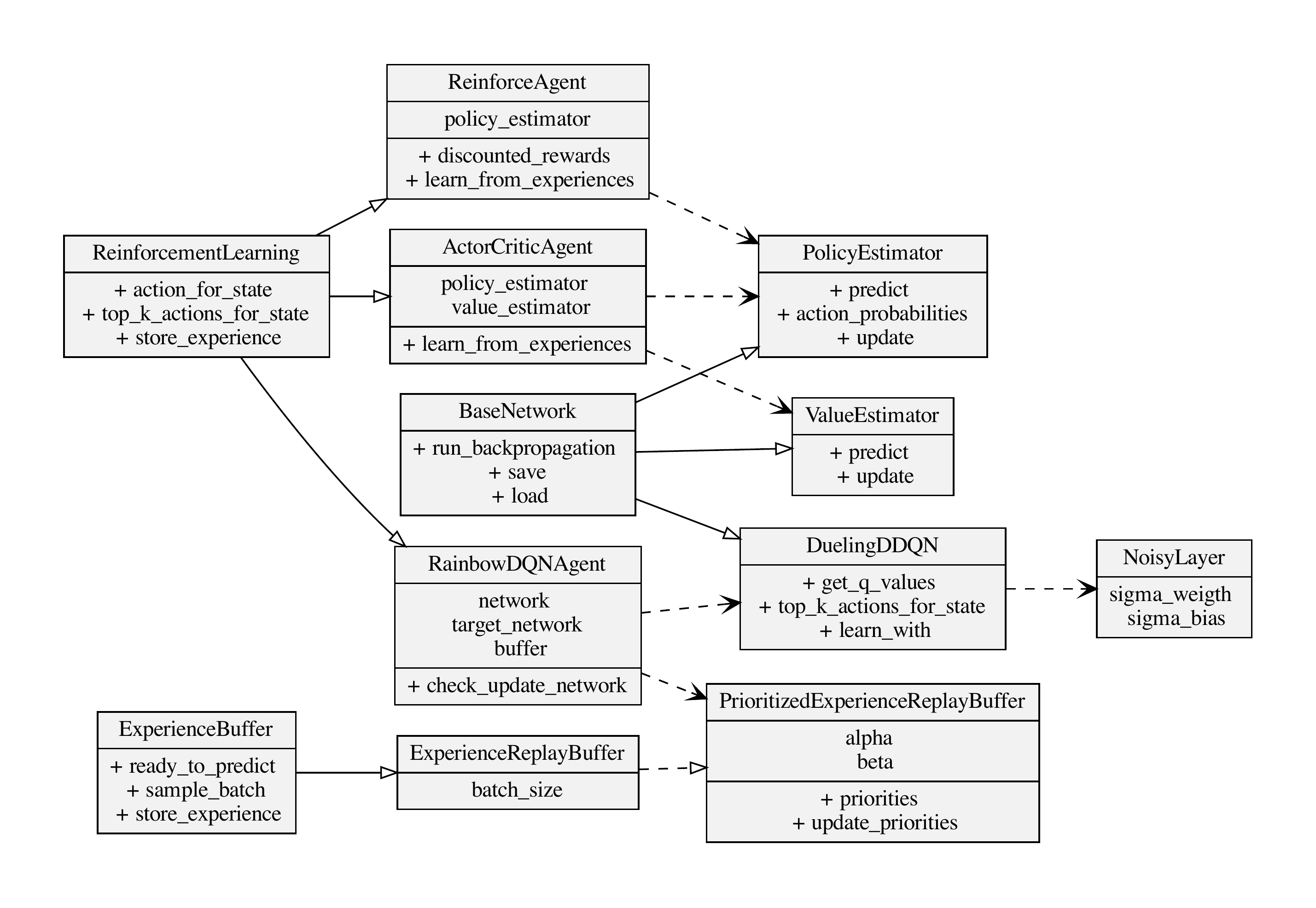}
	\caption{Class diagram covering the most important parts of the implementation.}
	\label{fig:classdiagram}
\end{figure}

The source code was designed using the \textit{Object Oriented Programming} (OOP) paradigm. Using inheritance to create component abstractions allowed for quicker development with less need for duplication. A class diagram containing the main classes, attributes, and functions is available in \cref{fig:classdiagram}. The main base classes created encapsulate common aspects of environments, neural networks, \acrshort{rl} agents, and the experience buffers.

The {\tt ReinforcementLearning} abstract class defines what functions agents are expected to have. Namely, to predict the next action to be taken on a given state, the next $k$ best actions (for slate environments), and to store a particular experience (i.e., a state, action, reward, done, next observed state) tuple. The {\tt BaseNetwork} base class implements helper functions like saving and loading from a file, freezing specific parameters, running backward propagation of a loss, plotting the gradient graph, configuring the hardware device (i.e., CPU or GPU). The {\tt ExperienceBuffer} interface expects implementations for experiences to be stored, sampled, and for the buffer to tell if there are enough experiences to start making predictions. Additionally, a {\tt LearningStatistics} module was developed to help collect different metrics that agents may output while training, providing ways to retrieve, plot, and aggregate them on many levels (i.e., model, episode, time-step, environment). Finally, the {\tt Manager} module coordinates the sending and receiving of actions and states. Managers help with training agents, hyperparameter search, executing episodes, and printing overviews of environments.

The project was developed using modern best-practices, such as automatic code formatting and linting (using python packages \textit{black} and \textit{flake8}) that help with the readability of the code. The project also uses a dependency-management tool to download, version, and update the several libraries needed to run the project (using the \textit{poetry} package), which makes the project usable with just a few commands. Unit tests and coverage measurement were implemented to help provide a stable and reliable project (using \textit{pytest} and \textit{coverage}). Finally, it is possible to generate \textit{HTML} documentation of all the modules (using \textit{pdoc}). All these tools are easy to use, and described in the project's documentation.

\chapter{Results And Discussion}
\label{chapter:results}

\section{Preliminary Results}

\begin{figure}[!htbp]
	\centering
	\includegraphics[width=1\textwidth]{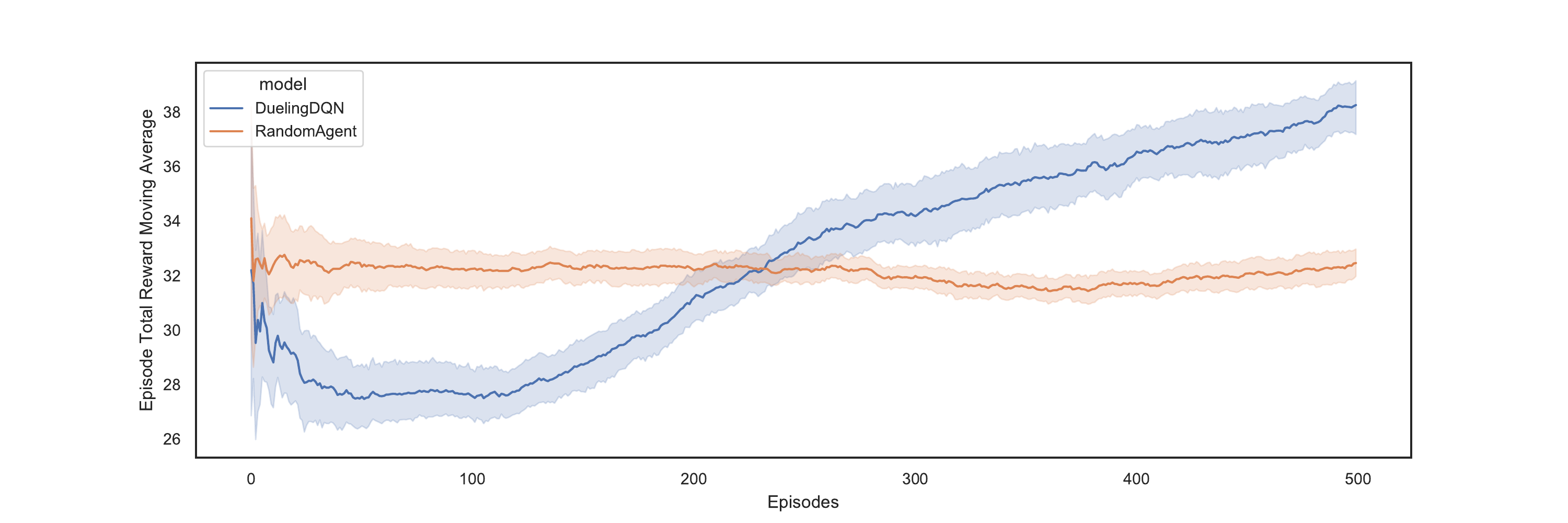}
	\caption{Comparison of an agent that recommends movies randomly (orange) and an agent that learns from user reviews (blue).}
	\label{fig:preliminaryresults}
\end{figure}

Using the already discussed environment, we can run a comparison of a random agent and a Dueling DQN agent (as shown in \cref{fig:dueling}), to see how they fare in terms of rewards obtained. For the comparison, agents learned for 500 episodes, and each episode contained 50 recommendations to a randomly selected user. This test was executed 20 times for each agent, and \cref{fig:preliminaryresults} shows the average moving reward of the previous 100 iterations, along with a confidence interval of $95\%$ confidence. It is possible to see, at the last episode's reward average, that the DQN agent had users rating around $3.8/5$. Meanwhile, the random agent had ratings around $3.2/5$. 

This is already a significant difference in terms of user satisfaction, and the plot indicates that further learning iterations would only increase the difference. Also, no hyperparameter optimization was done in this model. This preliminary result approximates the \citeyear{choi_reinforcement_2018} findings of \citet{choi_reinforcement_2018}, whose models (Q-Leaning and SARSA) approximate return 8 out of 10 after 200 episodes.

\section{Hyperparameter Search}

When comparing the results of the executions, it was found that taking only the means was not enough to make a decision on the optimal parameters. The standard deviation also provided valuable data to choose the best hyperparameters. To that end, the following plots contain the $95\%$ confidence interval of the results of the executions, along with the means. The complete execution results can be found in Appendix \ref{ap:paramsresults}.

\begin{figure}[!htbp]
	\centering
	\includegraphics[width=0.9\textwidth]{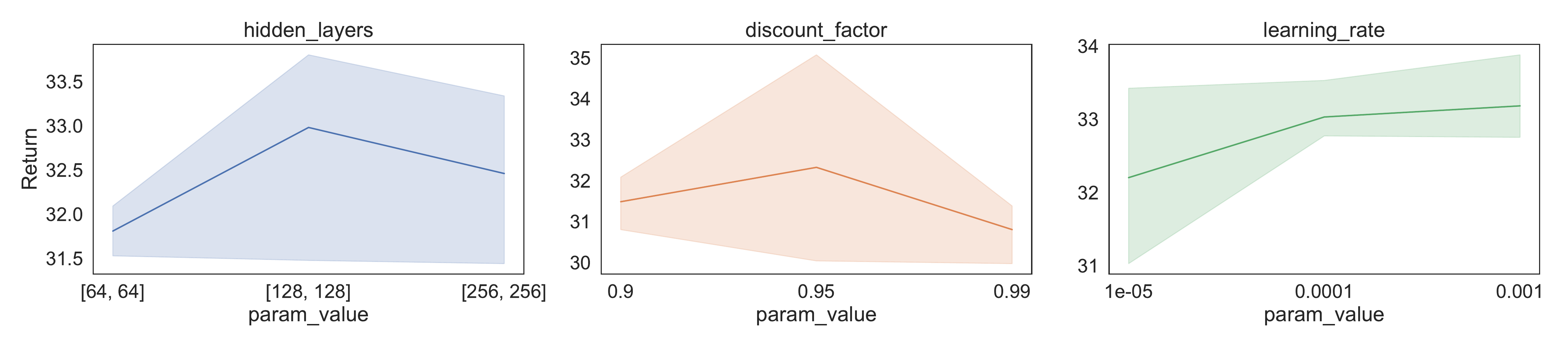}
	\caption{Average final moving reward out of 100 episodes, for different parameters of the REINFORCE agent. Notice how $\gamma=0.95$ had the highest average, but also the highest variance. Therefore it would be wiser to choose $0.9$ in this case.}
	\label{fig:paramsreinforce}
\end{figure}

The results for the hyperparameter search with the REINFORCE model can be found in \cref{fig:paramsreinforce}. We found that the policy estimator network with two hidden layers of 128 neurons worked best. Additionally, we found highest returns with a discount factor $\gamma$ of $90\%$ and a learning rate $\eta$ of 0.001.

\begin{figure}[!htbp]
	\centering
	\includegraphics[width=0.9\textwidth]{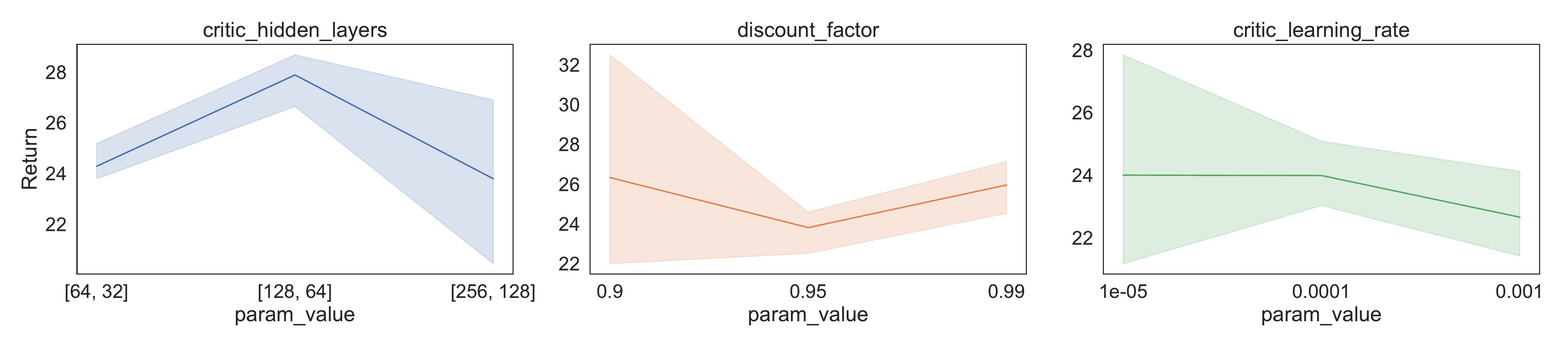}
	\caption{Average final moving reward out of 100 episodes, for different parameters of the Actor-Critic agent.}
	\label{fig:paramsac}
\end{figure}

Since the policy estimator is similar to the Actor network in the Actor-Critic agent, we focused its tests on the critic parameters. Results can be found in \cref{fig:paramsac}. The best critic network had hidden layers of 128 and 64 neurons, a $\gamma$ of $0.99$, and $\eta_{critic}$ of $0.0001$.

\begin{figure}[!htbp]
	\centering
	\includegraphics[width=1\textwidth]{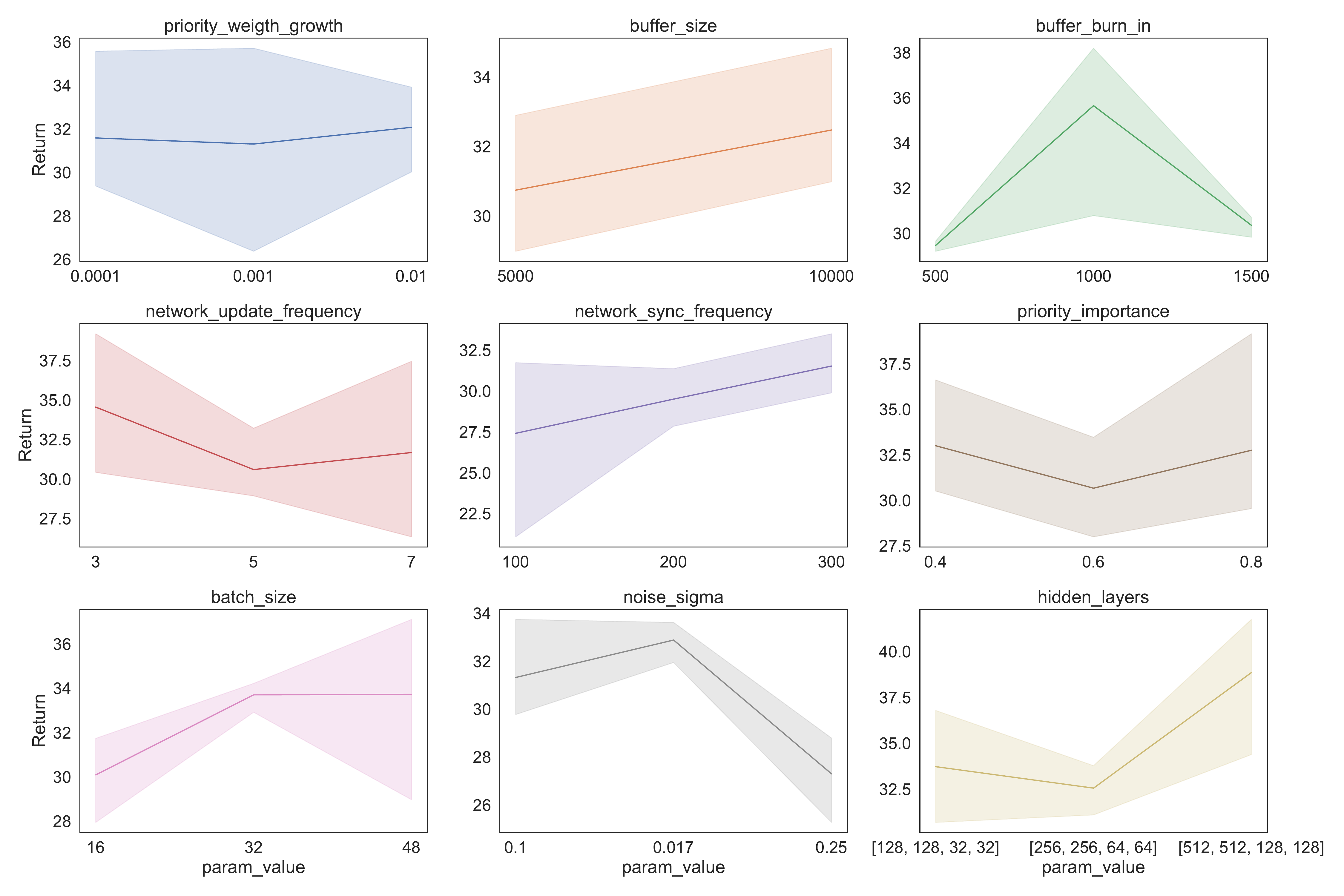}
	\caption{Average final moving reward out of 100 episodes, for different parameters of the Dueling DQN agent.}
	\label{fig:paramsdqn}
\end{figure}

The results from the Dueling DQN parameter search can be found in \cref{fig:paramsdqn}. It was found that the best network update frequency was 3, the target network sync frequency was $300$, the priority importance was $0.4$, the priority weight growth ($\beta$-annealing) was $0.01$, the buffer size was 10000, the buffer burn-in was $1000$, the batch size was $32$, the gaussian noise $\sigma$ was $0.017$, and the best architecture was with two hidden layers of $512$ neurons, and $128$ neurons for the advantage and value sub-networks. During the tests we used a $\gamma$ of $0.95$, and $\eta$ of $0.001$.

\section{Benchmark Results}

\begin{figure}[!htbp]
	\centering
	\includegraphics[width=1\textwidth]{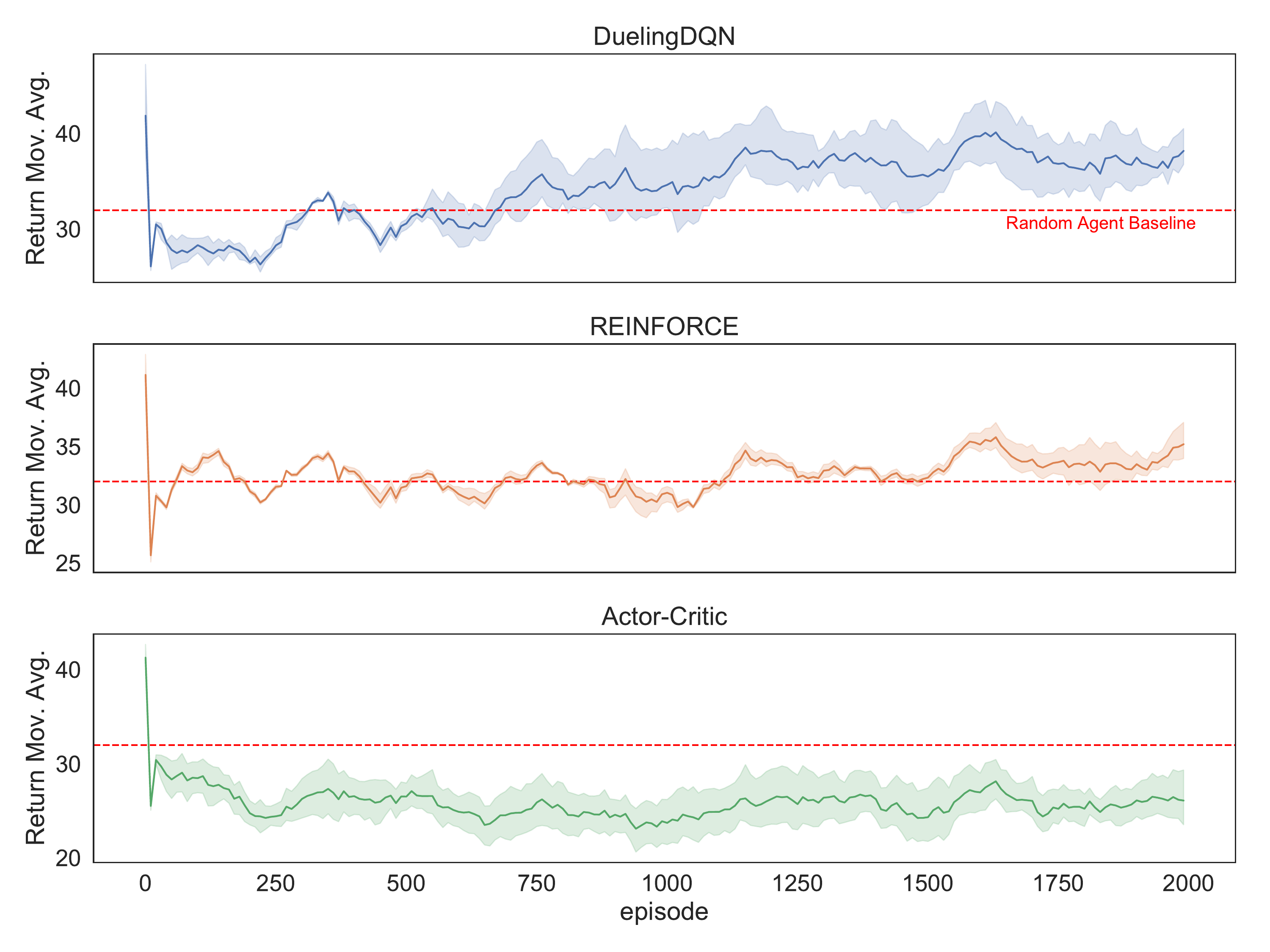}
	\caption{Comparison of training of the agents with the best hyper-parameters found. The horizontal line at return $32$, symbolizing the performance of a random agent, was added to help compare them. The area around the lines is the $95\%$ confidence interval. The Dueling DQN agent was trained for 3 seeds, while the others were trained for 5. In this scenario, the maximum possible reward is 50, and it would mean that the user rated 5/5 for every single recommended movie.}
	\label{fig:combined_execution_results}
\end{figure}

The results of the benchmark for all the agents can be found in \cref{fig:combined_execution_results}. The Dueling DQN agent was trained for $3$ seeds, while the others were trained for $5$. Inspecting the plot allows to identify that the DQN agent clearly outperformed the others. Moreover, it seems that the Actor-Critic agent hasn't managed to get results better than random. There may be some extra challenges in modeling the recommendation choices as probability distributions. Given that those agents have stochastic policies, it may be that small differences in the output nodes cause sub-optimal movies to be sampled too often. It may also be that it needed more training episodes, due to the high variance of the reward signal. A combination of more sophisticated Actor-Critic agents (e.g., \textit{Soft Actor-Critic}, \acrshort{a2c}) could potentially bring powerful improvements as well.

An interesting factor to analyze in this sort of setup is that the agent starts off without any previous knowledge of the environment. In the \acrshort{recsys} literature this is referred to as the \gls{coldstartproblem}. The ability of an agent to deliver as little ``bad'' recommendations as possible is very valuable in commercial situations, when the stakes are higher. From this perspective, and reviewing the results from \cref{fig:combined_execution_results}, the REINFORCE agent seems to make less errors in the short term. In the long term, however, it fails to deliver  predictions as good as the ones from the DQN agent.

\section{Relevance Measurements}

Using the best model found (the DQN agent), we can execute simulations with slates (i.e., when an agent recommends multiple movies at each time-step). This allows us to calculate more traditional \acrshort{recsys} relevance metrics. Results can be found in \cref{table:relevance_metrics_describe}.

\begin{table}[!htbp]
\centering
\begin{tabular}{l}
\begin{filecontents*}{csv/relevance_metrics_describe.csv}
,DCG@10 Random,DCG@10 DQN,NDCG@10 Random,NDCG@10 DQN
Mean,17.5560,19.5429,0.7727,0.8602
Std,4.5842,2.2787,0.2017,0.1003
Min,9.2098,13.1055,0.4054,0.5768
Median,19.2187,20.1940,0.8459,0.8889
Max,21.5615,22.7177,0.9491,1.0
\end{filecontents*}
\csvautotabular{csv/relevance_metrics_describe.csv}
\end{tabular}
\caption{Comparison of relevance metrics for an agent making random recommendations and a Dueling DQN agent.}
\label{table:relevance_metrics_describe}
\end{table}

It is important to note that the results from other researchers mentioned before had different ways to model relevance. While in their case a relevant movie is a movie that was reviewed by a user, in our case relevance is the simulated rating (from 1 to 5). Additionally, when calculating the ideal relevance, we used ratings of 5 (i.e., an ideal slate is one in which the chosen recommended movie receives a 5/5 review). Since our agent maximizes the simulated rating (part of our definition of reward), it was expected that the agent would perform better than random. 

If we measured other interesting \acrshort{recsys} metrics, like the diversity of recommended movie genres, it is possible that we would not see great results. This sort of investigation, optimizing for many objectives, is beyond the scope of this work.

\chapter{Conclusions}
\label{chapter:conclusions}

A comprehensive search was performed to find proper \acrshort{rl} environments that simulate realistic problems of a \acrlong{recsys}. Having found a most suitable one, improvements were made to fix the problems it had. Agents using different learning strategies were tested in this novel environment, and the results were comparable to those found by other recent studies. The code was made available and open-source at Github\footnote{\href{https://github.com/luksfarris/pydeeprecsys}{github.com/luksfarris/pydeeprecsys}}.

The usage of \acrshort{rl} allowed for an effective build of a hybrid \acrshort{recsys} approach, that combines features of model-based \gls{collaborativefiltering},  and \gls{casebased} systems. It is shown how even though the learning models start with zero information about the items and users, some are able to provide meaningful recommendations after around $200$ simulated user sessions, thus effectively handling the \gls{coldstartproblem}.

Three different \acrshort{rl} models were implemented, and some hyperparameter search was conducted to configure them as well as possible. It was found that the model based on estimating Q-values using \acrshort{dl} worked better consistently. Particularly, the models that had stochastic policies did not perform as well. 

Results were compared based on the final episode reward sum average, to demonstrate that models effectively outperform random recommendation strategies. The \acrshort{ndcg}@10 metric was computed for the \acrshort{dqn}-based model, to demonstrate that it effectively returns relevant recommendation slates.

\chapter{Suggestions for Future Research}
\label{chapter:suggestions}

One of the biggest challenges of researching this topic was finding suitable learning environments. Open-source environments were scarce, not actively maintained, only executable on specific Operational Systems (i.e., Linux), only executable on specific Python versions (i.e., 3.6), and poorly documented. This represents a technical barrier, that prevents the field from advancing faster. A great example of learning environments that work out-of-the-box on most operational systems and python versions is Open AI Gym, and it should serve as an inspiration for more \acrshort{recsys} environments. Additionally, more heterogeneous and larger datasets could be included, such as the \textit{MovieLens1B} dataset or the ones from the ACM conferences.

A second important branch of investigation is on the simulation of users. In other words, given a ground-truth dataset of user-journeys, finding a strategy that optimally generalizes how users behave in those particular conditions. While matrix factorization strategies are fast, practical, and relatively accurate, their performance might not scale as well as neural-network based approaches. In particular, it seems that Recurrent Neural Networks would be a powerful candidate in this particular problem.

A third interesting, and maybe the hardest, problem suitable for future research is finding appropriate multi-objective rewards, that simultaneously keep users satisfied with the recommendations, but without compromising the user, nor the items being recommended. For instance, an optimized food recommender system could recommend only chocolates to a user if health factors are not considered in the reward. Alternatively, an optimized clothing recommender system could recommend only pieces from the three major stores, effectively killing the competition. Encoding counterfactual elements in the reward signal could prevent such problems.

A fourth, very interesting path of research is regarding the \acrlong{rl} agents. More sophisticated Actor-Critic methods seem to have been gaining traction recently on other applications, such as games and robotics. In particular, it would be interesting to understand if \acrlong{a2c} or Soft-Actor-Critic (SAC) agents could outperform the Dueling DQN model, both with deterministic and stochastic policies. A more thorough comparison would also be very interesting, checking if tabular methods would be feasible, and if their performance would be comparable.

Lastly, there is a lot of room for research regarding the explainability of recommendations. Particularly in \acrshort{recsys} environments, the power of being able to explain why certain recommendations are being made is invaluable. While explainability in supervised learning problems has been explored extensively, there is still a gap in the Deep Reinforcement Learning application field. Specifically, users could be interested in what is the expected relevance of a particular document, what factors were relevant in estimating it, and how each factor weighs in.

\addcontentsline{toc}{chapter}{Bibliography}
\bibliographystyle{unsrtnat}
\bibliography{referencias}

\begin{thebibliography}{52}
\providecommand{\natexlab}[1]{#1}
\providecommand{\url}[1]{\texttt{#1}}
\expandafter\ifx\csname urlstyle\endcsname\relax
  \providecommand{\doi}[1]{doi: #1}\else
  \providecommand{\doi}{doi: \begingroup \urlstyle{rm}\Url}\fi

\bibitem[Gospodnetic et~al.(2010)Gospodnetic, Hatcher, and McCandless]{lucene}
Otis Gospodnetic, Erik Hatcher, and Michael McCandless.
\newblock \emph{Lucene in {Action}}.
\newblock Manning Publications, 2nd edition, 2010.

\bibitem[Falk(2019)]{falk_practical_2019}
Kim Falk.
\newblock \emph{Practical {Recommender} {Systems}}.
\newblock Manning Publications, first edition, January 2019.
\newblock ISBN 978-1-61729-270-5.
\newblock URL
  \url{https://learning.oreilly.com/library/view/practical-recommender-systems/9781617292705/}.
\newblock Accessed on 2021-02-27.

\bibitem[Kaptein(2015)]{persuasion}
Maurits Kaptein.
\newblock \emph{Persuasion Profiling}.
\newblock Business Contact Publishers, 2015.

\bibitem[Santana et~al.(2020{\natexlab{a}})Santana, Melo, Camargo, Brandão,
  Soares, Oliveira, and Caetano]{mars}
Marlesson R.~O. Santana, Luckeciano~C. Melo, Fernando H.~F. Camargo, Bruno
  Brandão, Anderson Soares, Renan~M. Oliveira, and Sandor Caetano.
\newblock {MARS}-{Gym}: {A} {Gym} framework to model, train, and evaluate
  {Recommender} {Systems} for {Marketplaces}.
\newblock \emph{arXiv:2010.07035 [cs, stat]}, September 2020{\natexlab{a}}.

\bibitem[cha(2019)]{challenge}
\emph{RecSys Challenge '19: Proceedings of the Workshop on ACM Recommender
  Systems Challenge}, New York, NY, USA, 2019. Association for Computing
  Machinery.
\newblock ISBN 9781450376679.

\bibitem[Ge et~al.(2010)Ge, Delgado, and Jannach]{recsysmetrics}
Mouzhi Ge, Carla Delgado, and Dietmar Jannach.
\newblock Beyond accuracy: {Evaluating} recommender systems by coverage and
  serendipity.
\newblock In \emph{Proceedings of the Fourth ACM Conference on Recommender
  Systems}, pages 257--260, January 2010.
\newblock \doi{10.1145/1864708.1864761}.

\bibitem[Chapman et~al.(2000)Chapman, Clinton, Kerber, Khabaza, Reinartz,
  Shearer, and Wirth]{crispdm}
Pete Chapman, Julian Clinton, Randy Kerber, Thomas Khabaza, Thomas Reinartz,
  Colin Shearer, and Rudiger Wirth.
\newblock Crisp-dm 1.0 step-by-step data mining guide.
\newblock Technical report, The CRISP-DM consortium, August 2000.

\bibitem[Silyn-Roberts(2000)]{writingforscience}
Heather Silyn-Roberts.
\newblock \emph{Writing {For} {Science} {And} {Engineering}}.
\newblock Elsevier, 2000.
\newblock ISBN 978-0-7506-4636-9.
\newblock URL \url{https://doi.org/10.1016/B978-0-7506-4636-9.X5000-9}.
\newblock Accessed on 2021-03-13.

\bibitem[Aggarwal(2016)]{recommendersystems}
Charu~C. Aggarwal.
\newblock \emph{Recommender {Systems}}.
\newblock Springer International Publishing, Cham, 2016.
\newblock ISBN 978-3-319-29657-9 978-3-319-29659-3.
\newblock \doi{10.1007/978-3-319-29659-3}.
\newblock URL \url{http://link.springer.com/10.1007/978-3-319-29659-3}.
\newblock Accessed on 2021-03-13.

\bibitem[Sutton and Barto(2018)]{reinforcementlearning}
Richard~S. Sutton and Andrew~G. Barto.
\newblock \emph{Reinforcement learning: an introduction}.
\newblock Adaptive computation and machine learning. MIT Press, Cambridge,
  Mass, 2 edition, 2018.
\newblock ISBN 978-0-262-19398-6.

\bibitem[Goodfellow et~al.(2016)Goodfellow, Bengio, and
  Courville]{deeplearning}
Ian Goodfellow, Yoshua Bengio, and Aaron Courville.
\newblock \emph{Deep {Learning}}.
\newblock MIT Press, 2016.
\newblock URL \url{http://www.deeplearningbook.org}.
\newblock Accessed on 2021-03-03.

\bibitem[Leskovec et~al.(2019)Leskovec, Rajaraman, and
  Ullman]{leskovec_mining_2019}
Jure Leskovec, Anand Rajaraman, and Jeff Ullman.
\newblock \emph{Mining {Of} {Massive} {Datasets}}.
\newblock Cambridge University Press, 3 edition, 2019.
\newblock URL \url{http://www.mmds.org}.
\newblock Accessed on 2021-03-03.

\bibitem[Schaul et~al.(2016)Schaul, Quan, Antonoglou, and
  Silver]{schaul_prioritized_2016}
Tom Schaul, John Quan, Ioannis Antonoglou, and David Silver.
\newblock Prioritized {Experience} {Replay}.
\newblock \emph{arXiv:1511.05952 [cs]}, February 2016.
\newblock URL \url{http://arxiv.org/abs/1511.05952}.
\newblock arXiv: 1511.05952. Accessed on 2021-03-18.

\bibitem[Hasselt et~al.(2016)Hasselt, Guez, and Silver]{hasselt_deep_2016}
Hado~van Hasselt, Arthur Guez, and David Silver.
\newblock Deep {Reinforcement} {Learning} with {Double} {Q}-{Learning}.
\newblock \emph{Proceedings of the AAAI Conference on Artificial Intelligence},
  30\penalty0 (1), March 2016.
\newblock ISSN 2374-3468.
\newblock URL \url{https://ojs.aaai.org/index.php/AAAI/article/view/10295}.
\newblock Number: 1. Accessed on 2021-03-18.

\bibitem[Fortunato et~al.(2019)Fortunato, Azar, Piot, Menick, Osband, Graves,
  Mnih, Munos, Hassabis, Pietquin, Blundell, and Legg]{fortunato_noisy_2019}
Meire Fortunato, Mohammad~Gheshlaghi Azar, Bilal Piot, Jacob Menick, Ian
  Osband, Alex Graves, Vlad Mnih, Remi Munos, Demis Hassabis, Olivier Pietquin,
  Charles Blundell, and Shane Legg.
\newblock Noisy {Networks} for {Exploration}.
\newblock \emph{arXiv:1706.10295 [cs, stat]}, July 2019.
\newblock URL \url{http://arxiv.org/abs/1706.10295}.
\newblock arXiv: 1706.10295. Accessed on 2021-03-18.

\bibitem[Wang et~al.(2016)Wang, Schaul, Hessel, Hasselt, Lanctot, and
  Freitas]{wang_dueling_2016}
Ziyu Wang, Tom Schaul, Matteo Hessel, Hado Hasselt, Marc Lanctot, and Nando
  Freitas.
\newblock Dueling {Network} {Architectures} for {Deep} {Reinforcement}
  {Learning}.
\newblock In \emph{International {Conference} on {Machine} {Learning}}, pages
  1995--2003. PMLR, June 2016.
\newblock URL \url{http://proceedings.mlr.press/v48/wangf16.html}.
\newblock ISSN: 1938-7228. Accessed on 2021-03-18.

\bibitem[Silver et~al.(2014)Silver, Lever, Heess, Degris, Wierstra, and
  Riedmiller]{silver_deterministic_2014}
David Silver, Guy Lever, Nicolas Heess, Thomas Degris, Daan Wierstra, and
  Martin Riedmiller.
\newblock Deterministic {Policy} {Gradient} {Algorithms}.
\newblock In \emph{International {Conference} on {Machine} {Learning}}, pages
  387--395. PMLR, January 2014.
\newblock URL \url{http://proceedings.mlr.press/v32/silver14.html}.
\newblock ISSN: 1938-7228. Accessed on 2021-03-18.

\bibitem[Mnih et~al.(2016)Mnih, Badia, Mirza, Graves, Lillicrap, Harley,
  Silver, and Kavukcuoglu]{mnih_asynchronous_2016}
Volodymyr Mnih, Adria~Puigdomenech Badia, Mehdi Mirza, Alex Graves, Timothy
  Lillicrap, Tim Harley, David Silver, and Koray Kavukcuoglu.
\newblock Asynchronous {Methods} for {Deep} {Reinforcement} {Learning}.
\newblock In \emph{International {Conference} on {Machine} {Learning}}, pages
  1928--1937. PMLR, June 2016.
\newblock URL \url{http://proceedings.mlr.press/v48/mniha16.html}.
\newblock ISSN: 1938-7228. Accessed on 2021-03-18.

\bibitem[Zhang et~al.(2019)Zhang, Yao, Sun, and Tay]{deeplearningrecsysreview}
Shuai Zhang, Lina Yao, Aixin Sun, and Yi~Tay.
\newblock Deep {Learning} {Based} {Recommender} {System}: {A} {Survey} and
  {New} {Perspectives}.
\newblock \emph{ACM Computing Surveys}, 52\penalty0 (1):\penalty0 5:1--5:38,
  February 2019.
\newblock ISSN 0360-0300.
\newblock \doi{10.1145/3285029}.
\newblock URL \url{https://doi.org/10.1145/3285029}.
\newblock Accessed on 2021-03-15.

\bibitem[Gupta and Katarya(2018)]{gupta_study_2018}
G.~Gupta and R.~Katarya.
\newblock A {Study} of {Recommender} {Systems} {Using} {Markov} {Decision}
  {Process}.
\newblock In \emph{2018 {Second} {International} {Conference} on {Intelligent}
  {Computing} and {Control} {Systems} ({ICICCS})}, pages 1279--1283, June 2018.
\newblock \doi{10.1109/ICCONS.2018.8663161}.

\bibitem[Mahmood and Ricci(2007)]{mahmood_learning_2007}
Tariq Mahmood and Francesco Ricci.
\newblock Learning and adaptivity in interactive recommender systems.
\newblock In \emph{Proceedings of the ninth international conference on
  {Electronic} commerce}, {ICEC} '07, pages 75--84, New York, NY, USA, August
  2007. Association for Computing Machinery.
\newblock ISBN 978-1-59593-700-1.
\newblock \doi{10.1145/1282100.1282114}.
\newblock URL \url{https://doi.org/10.1145/1282100.1282114}.
\newblock Accessed on 2021-03-13.

\bibitem[Mahmood and Ricci(2009)]{mahmood_improving_2009}
Tariq Mahmood and Francesco Ricci.
\newblock Improving recommender systems with adaptive conversational
  strategies.
\newblock In \emph{Proceedings of the 20th {ACM} conference on {Hypertext} and
  hypermedia}, {HT} '09, pages 73--82, New York, NY, USA, June 2009.
  Association for Computing Machinery.
\newblock ISBN 978-1-60558-486-7.
\newblock \doi{10.1145/1557914.1557930}.
\newblock URL \url{https://doi.org/10.1145/1557914.1557930}.
\newblock Accessed on 2021-03-14.

\bibitem[Li et~al.(2010)Li, Chu, Langford, and
  Schapire]{li_contextual-bandit_2010}
Lihong Li, Wei Chu, John Langford, and Robert~E. Schapire.
\newblock A contextual-bandit approach to personalized news article
  recommendation.
\newblock In \emph{Proceedings of the 19th international conference on {World}
  wide web}, {WWW} '10, pages 661--670, New York, NY, USA, April 2010.
  Association for Computing Machinery.
\newblock ISBN 978-1-60558-799-8.
\newblock \doi{10.1145/1772690.1772758}.
\newblock URL \url{https://doi.org/10.1145/1772690.1772758}.
\newblock Accessed on 2021-03-13.

\bibitem[Li et~al.(2011)Li, Chu, Langford, and Wang]{li_unbiased_2011}
Lihong Li, Wei Chu, John Langford, and Xuanhui Wang.
\newblock Unbiased offline evaluation of contextual-bandit-based news article
  recommendation algorithms.
\newblock In \emph{Proceedings of the fourth {ACM} international conference on
  {Web} search and data mining}, {WSDM} '11, pages 297--306, New York, NY, USA,
  February 2011. Association for Computing Machinery.
\newblock ISBN 978-1-4503-0493-1.
\newblock \doi{10.1145/1935826.1935878}.
\newblock URL \url{https://doi.org/10.1145/1935826.1935878}.
\newblock Accessed on 2021-03-14.

\bibitem[Kale et~al.(2010)Kale, Reyzin, and Schapire]{kale_non-stochastic_2010}
Satyen Kale, Lev Reyzin, and Robert~E. Schapire.
\newblock Non-stochastic bandit slate problems.
\newblock In \emph{Proceedings of the 23rd {International} {Conference} on
  {Neural} {Information} {Processing} {Systems} - {Volume} 1}, {NIPS}'10, pages
  1054--1062, Red Hook, NY, USA, December 2010. Curran Associates Inc.

\bibitem[Chu et~al.(2011)Chu, Li, Reyzin, and Schapire]{chu_contextual_2011}
Wei Chu, Lihong Li, Lev Reyzin, and Robert Schapire.
\newblock Contextual {Bandits} with {Linear} {Payoff} {Functions}.
\newblock In \emph{Proceedings of the {Fourteenth} {International} {Conference}
  on {Artificial} {Intelligence} and {Statistics}}, pages 208--214. JMLR
  Workshop and Conference Proceedings, June 2011.
\newblock URL \url{http://proceedings.mlr.press/v15/chu11a.html}.
\newblock ISSN: 1938-7228. Accessed on 2021-03-14.

\bibitem[Bouneffouf et~al.(2012)Bouneffouf, Bouzeghoub, and
  Gançarski]{bouneffouf_contextual-bandit_2012}
Djallel Bouneffouf, Amel Bouzeghoub, and Alda~Lopes Gançarski.
\newblock A {Contextual}-{Bandit} {Algorithm} for {Mobile} {Context}-{Aware}
  {Recommender} {System}.
\newblock In Tingwen Huang, Zhigang Zeng, Chuandong Li, and Chi~Sing Leung,
  editors, \emph{Neural {Information} {Processing}}, Lecture {Notes} in
  {Computer} {Science}, pages 324--331, Berlin, Heidelberg, 2012. Springer.
\newblock ISBN 978-3-642-34487-9.
\newblock \doi{10.1007/978-3-642-34487-9_40}.

\bibitem[Thomas et~al.(2015)Thomas, Theocharous, and
  Ghavamzadeh]{thomas_high-confidence_2015}
Philip Thomas, Georgios Theocharous, and Mohammad Ghavamzadeh.
\newblock High-{Confidence} {Off}-{Policy} {Evaluation}.
\newblock \emph{Proceedings of the AAAI Conference on Artificial Intelligence},
  29\penalty0 (1), February 2015.
\newblock ISSN 2374-3468.
\newblock URL \url{https://ojs.aaai.org/index.php/AAAI/article/view/9541}.
\newblock Number: 1. Accessed on 2021-03-13.

\bibitem[Theocharous et~al.(2015)Theocharous, Thomas, and
  Ghavamzadeh]{theocharous_ad_2015}
Georgios Theocharous, Philip~S. Thomas, and Mohammad Ghavamzadeh.
\newblock Ad {Recommendation} {Systems} for {Life}-{Time} {Value}
  {Optimization}.
\newblock In \emph{Proceedings of the 24th {International} {Conference} on
  {World} {Wide} {Web}}, {WWW} '15 {Companion}, pages 1305--1310, New York, NY,
  USA, May 2015. Association for Computing Machinery.
\newblock ISBN 978-1-4503-3473-0.
\newblock \doi{10.1145/2740908.2741998}.
\newblock URL \url{https://doi.org/10.1145/2740908.2741998}.
\newblock Accessed on 2021-03-13.

\bibitem[Choi et~al.(2018)Choi, Ha, Hwang, Kim, Ha, and
  Yoon]{choi_reinforcement_2018}
Sungwoon Choi, Heonseok Ha, Uiwon Hwang, Chanju Kim, Jung-Woo Ha, and Sungroh
  Yoon.
\newblock Reinforcement {Learning} based {Recommender} {System} using
  {Biclustering} {Technique}.
\newblock January 2018.
\newblock URL \url{https://arxiv.org/abs/1801.05532v1}.
\newblock Accessed on 2021-03-16.

\bibitem[Zheng et~al.(2018)Zheng, Zhang, Zheng, Xiang, Yuan, Xie, and
  Li]{zheng_drn_2018}
Guanjie Zheng, Fuzheng Zhang, Zihan Zheng, Yang Xiang, Nicholas~Jing Yuan, Xing
  Xie, and Zhenhui Li.
\newblock {DRN}: {A} {Deep} {Reinforcement} {Learning} {Framework} for {News}
  {Recommendation}.
\newblock In \emph{Proceedings of the 2018 {World} {Wide} {Web} {Conference}},
  {WWW} '18, pages 167--176, Republic and Canton of Geneva, CHE, April 2018.
  International World Wide Web Conferences Steering Committee.
\newblock ISBN 978-1-4503-5639-8.
\newblock \doi{10.1145/3178876.3185994}.
\newblock URL \url{https://doi.org/10.1145/3178876.3185994}.
\newblock Accessed on 2021-03-15.

\bibitem[Chen et~al.(2018)Chen, Yu, Da, Tan, Huang, and
  Tang]{chen_stabilizing_2018}
Shi-Yong Chen, Yang Yu, Qing Da, Jun Tan, Hai-Kuan Huang, and Hai-Hong Tang.
\newblock Stabilizing {Reinforcement} {Learning} in {Dynamic} {Environment}
  with {Application} to {Online} {Recommendation}.
\newblock In \emph{Proceedings of the 24th {ACM} {SIGKDD} {International}
  {Conference} on {Knowledge} {Discovery} \& {Data} {Mining}}, {KDD} '18, pages
  1187--1196, New York, NY, USA, July 2018. Association for Computing
  Machinery.
\newblock ISBN 978-1-4503-5552-0.
\newblock \doi{10.1145/3219819.3220122}.
\newblock URL \url{https://doi.org/10.1145/3219819.3220122}.
\newblock Accessed on 2021-03-15.

\bibitem[Zhao et~al.(2018{\natexlab{a}})Zhao, Zhang, Ding, Xia, Tang, and
  Yin]{zhao_recommendations_2018}
Xiangyu Zhao, Liang Zhang, Zhuoye Ding, Long Xia, Jiliang Tang, and Dawei Yin.
\newblock Recommendations with {Negative} {Feedback} via {Pairwise} {Deep}
  {Reinforcement} {Learning}.
\newblock In \emph{Proceedings of the 24th {ACM} {SIGKDD} {International}
  {Conference} on {Knowledge} {Discovery} \& {Data} {Mining}}, {KDD} '18, pages
  1040--1048, New York, NY, USA, July 2018{\natexlab{a}}. Association for
  Computing Machinery.
\newblock ISBN 978-1-4503-5552-0.
\newblock \doi{10.1145/3219819.3219886}.
\newblock URL \url{https://doi.org/10.1145/3219819.3219886}.
\newblock Accessed on 2021-03-15.

\bibitem[Zhao et~al.(2018{\natexlab{b}})Zhao, Xia, Zhang, Ding, Yin, and
  Tang]{zhao_deep_2018}
Xiangyu Zhao, Long Xia, Liang Zhang, Zhuoye Ding, Dawei Yin, and Jiliang Tang.
\newblock Deep reinforcement learning for page-wise recommendations.
\newblock In \emph{Proceedings of the 12th {ACM} {Conference} on {Recommender}
  {Systems}}, {RecSys} '18, pages 95--103, New York, NY, USA, September
  2018{\natexlab{b}}. Association for Computing Machinery.
\newblock ISBN 978-1-4503-5901-6.
\newblock \doi{10.1145/3240323.3240374}.
\newblock URL \url{https://doi.org/10.1145/3240323.3240374}.
\newblock Accessed on 2021-03-09.

\bibitem[Brockman et~al.(2016)Brockman, Cheung, Pettersson, Schneider,
  Schulman, Tang, and Zaremba]{openaigym}
Greg Brockman, Vicki Cheung, Ludwig Pettersson, Jonas Schneider, John Schulman,
  Jie Tang, and Wojciech Zaremba.
\newblock Openai gym, 2016.

\bibitem[Ie et~al.(2019)Ie, Hsu, Mladenov, Jain, Narvekar, Wang, Wu, and
  Boutilier]{ie_recsim_2019}
Eugene Ie, Chih-wei Hsu, Martin Mladenov, Vihan Jain, Sanmit Narvekar, Jing
  Wang, Rui Wu, and Craig Boutilier.
\newblock {RecSim}: {A} {Configurable} {Simulation} {Platform} for
  {Recommender} {Systems}.
\newblock \emph{arXiv:1909.04847 [cs, stat]}, September 2019.
\newblock URL \url{http://arxiv.org/abs/1909.04847}.
\newblock arXiv: 1909.04847. Accessed on 2021-03-17.

\bibitem[Shi et~al.(2019)Shi, Ozsoy, Hurley, Smyth, Tragos, Geraci, and
  Lawlor]{shi_pyrecgym_2019}
Bichen Shi, Makbule~Gulcin Ozsoy, Neil Hurley, Barry Smyth, Elias~Z. Tragos,
  James Geraci, and Aonghus Lawlor.
\newblock {PyRecGym}: a reinforcement learning gym for recommender systems.
\newblock In \emph{Proceedings of the 13th {ACM} {Conference} on {Recommender}
  {Systems}}, {RecSys} '19, pages 491--495, New York, NY, USA, September 2019.
  Association for Computing Machinery.
\newblock ISBN 978-1-4503-6243-6.
\newblock \doi{10.1145/3298689.3346981}.
\newblock URL \url{https://doi.org/10.1145/3298689.3346981}.
\newblock Accessed on 2021-03-17.

\bibitem[Santana et~al.(2020{\natexlab{b}})Santana, Melo, Camargo, Brandão,
  Soares, Oliveira, and Caetano]{santana_mars-gym_2020}
Marlesson R.~O. Santana, Luckeciano~C. Melo, Fernando H.~F. Camargo, Bruno
  Brandão, Anderson Soares, Renan~M. Oliveira, and Sandor Caetano.
\newblock {MARS}-{Gym}: {A} {Gym} framework to model, train, and evaluate
  {Recommender} {Systems} for {Marketplaces}.
\newblock \emph{arXiv:2010.07035 [cs, stat]}, September 2020{\natexlab{b}}.
\newblock URL \url{http://arxiv.org/abs/2010.07035}.
\newblock arXiv: 2010.07035. Accessed on 2021-03-17.

\bibitem[Jannach et~al.(2020)Jannach, de~Souza P.~Moreira, and
  Oldridge]{jannach_why_2020}
Dietmar Jannach, Gabriel de~Souza P.~Moreira, and Even Oldridge.
\newblock Why {Are} {Deep} {Learning} {Models} {Not} {Consistently} {Winning}
  {Recommender} {Systems} {Competitions} {Yet}? {A} {Position} {Paper}.
\newblock In \emph{Proceedings of the {Recommender} {Systems} {Challenge}
  2020}, {RecSysChallenge} '20, pages 44--49, New York, NY, USA, September
  2020. Association for Computing Machinery.
\newblock ISBN 978-1-4503-8835-1.
\newblock \doi{10.1145/3415959.3416001}.
\newblock URL \url{https://doi.org/10.1145/3415959.3416001}.
\newblock Accessed on 2021-03-17.

\bibitem[Schifferer et~al.(2020)Schifferer, Titericz, Deotte, Henkel, Onodera,
  Liu, Tunguz, Oldridge, De~Souza Pereira~Moreira, and
  Erdem]{schifferer_gpu_2020}
Benedikt Schifferer, Gilberto Titericz, Chris Deotte, Christof Henkel, Kazuki
  Onodera, Jiwei Liu, Bojan Tunguz, Even Oldridge, Gabriel De~Souza
  Pereira~Moreira, and Ahmet Erdem.
\newblock {GPU} {Accelerated} {Feature} {Engineering} and {Training} for
  {Recommender} {Systems}.
\newblock In \emph{Proceedings of the {Recommender} {Systems} {Challenge}
  2020}, {RecSysChallenge} '20, pages 16--23, New York, NY, USA, September
  2020. Association for Computing Machinery.
\newblock ISBN 978-1-4503-8835-1.
\newblock \doi{10.1145/3415959.3415996}.
\newblock URL \url{https://doi.org/10.1145/3415959.3415996}.
\newblock Accessed on 2021-03-09.

\bibitem[D’Amour et~al.(2020)D’Amour, Srinivasan, Atwood, Baljekar,
  Sculley, and Halpern]{fairness_gym}
Alexander D’Amour, Hansa Srinivasan, James Atwood, Pallavi Baljekar,
  D.~Sculley, and Yoni Halpern.
\newblock Fairness is not static: Deeper understanding of long term fairness
  via simulation studies.
\newblock In \emph{Proceedings of the 2020 Conference on Fairness,
  Accountability, and Transparency}, FAccT ’20, page 525–534, New York, NY,
  USA, 2020. Association for Computing Machinery.
\newblock ISBN 9781450369367.
\newblock \doi{10.1145/3351095.3372878}.
\newblock URL \url{https://doi.org/10.1145/3351095.3372878}.
\newblock Accessed on 2021-03-18.

\bibitem[Harper and Konstan(2015)]{harper_movielens_2015}
F.~Maxwell Harper and Joseph~A. Konstan.
\newblock The {MovieLens} {Datasets}: {History} and {Context}.
\newblock \emph{ACM Transactions on Interactive Intelligent Systems},
  5\penalty0 (4):\penalty0 19:1--19:19, December 2015.
\newblock ISSN 2160-6455.
\newblock \doi{10.1145/2827872}.
\newblock URL \url{https://doi.org/10.1145/2827872}.
\newblock Accessed on 2021-05-05.

\bibitem[Vig et~al.(2012)Vig, Sen, and Riedl]{vig_tag_2012}
Jesse Vig, Shilad Sen, and John Riedl.
\newblock The {Tag} {Genome}: {Encoding} {Community} {Knowledge} to {Support}
  {Novel} {Interaction}.
\newblock \emph{ACM Transactions on Interactive Intelligent Systems},
  2\penalty0 (3):\penalty0 13:1--13:44, September 2012.
\newblock ISSN 2160-6455.
\newblock \doi{10.1145/2362394.2362395}.
\newblock URL \url{https://doi.org/10.1145/2362394.2362395}.
\newblock Accessed on 2021-05-05.

\bibitem[Koren et~al.(2009)Koren, Bell, and Volinsky]{koren_matrix_2009}
Yehuda Koren, Robert Bell, and Chris Volinsky.
\newblock Matrix {Factorization} {Techniques} for {Recommender} {Systems}.
\newblock \emph{Computer}, 42\penalty0 (8):\penalty0 30--37, August 2009.
\newblock ISSN 1558-0814.
\newblock \doi{10.1109/MC.2009.263}.
\newblock Conference Name: Computer.

\bibitem[Afkhamizadeh et~al.(2017)Afkhamizadeh, Avakov, and
  Takapoui]{afkhamizadeh_automated_nodate}
Mostafa Afkhamizadeh, Alexei Avakov, and Reza Takapoui.
\newblock Automated {Recommendation} {Systems}.
\newblock page~5, 2017.

\bibitem[Huang et~al.(2021)Huang, Fu, Li, Qu, Liu, and Chen]{huang_deep_2021}
Liwei Huang, Mingsheng Fu, Fan Li, Hong Qu, Yangjun Liu, and Wenyu Chen.
\newblock A deep reinforcement learning based long-term recommender system.
\newblock \emph{Knowledge-Based Systems}, 213:\penalty0 106706, February 2021.
\newblock ISSN 09507051.
\newblock \doi{10.1016/j.knosys.2020.106706}.
\newblock URL
  \url{https://linkinghub.elsevier.com/retrieve/pii/S0950705120308352}.
\newblock Accessed on 2021-03-09.

\bibitem[Sanz-Cruzado et~al.(2019)Sanz-Cruzado, Castells, and
  López]{sanz-cruzado_simple_2019}
Javier Sanz-Cruzado, Pablo Castells, and Esther López.
\newblock A simple multi-armed nearest-neighbor bandit for interactive
  recommendation.
\newblock In \emph{Proceedings of the 13th {ACM} {Conference} on {Recommender}
  {Systems}}, pages 358--362, Copenhagen Denmark, September 2019. ACM.
\newblock ISBN 978-1-4503-6243-6.
\newblock \doi{10.1145/3298689.3347040}.
\newblock URL \url{https://dl.acm.org/doi/10.1145/3298689.3347040}.
\newblock Accessed on 2021-03-09.

\bibitem[Zou et~al.(2019)Zou, Xia, Ding, Yin, Song, and
  Liu]{zou_reinforcement_2019}
Lixin Zou, Long Xia, Zhuoye Ding, Dawei Yin, Jiaxing Song, and Weidong Liu.
\newblock Reinforcement {Learning} to {Diversify} {Top}-{N} {Recommendation}.
\newblock In Guoliang Li, Jun Yang, Joao Gama, Juggapong Natwichai, and Yongxin
  Tong, editors, \emph{Database {Systems} for {Advanced} {Applications}},
  Lecture {Notes} in {Computer} {Science}, pages 104--120, Cham, 2019. Springer
  International Publishing.
\newblock ISBN 978-3-030-18579-4.
\newblock \doi{10.1007/978-3-030-18579-4_7}.

\bibitem[Yu et~al.(2020)Yu, Gu, Tao, Ge, and Chang]{yu_interactive_2020}
Yang Yu, Zhenhao Gu, Rong Tao, Jingtian Ge, and Kenglun Chang.
\newblock Interactive {Search} {Based} on {Deep} {Reinforcement} {Learning}.
\newblock \emph{arXiv:2012.06052 [cs]}, December 2020.
\newblock URL \url{http://arxiv.org/abs/2012.06052}.
\newblock arXiv: 2012.06052. Accessed on 2021-02-22.

\bibitem[Liu et~al.(2020{\natexlab{a}})Liu, Tang, Guo, Li, Ye, and
  He]{liu_top-aware_2020}
Feng Liu, Ruiming Tang, Huifeng Guo, Xutao Li, Yunming Ye, and Xiuqiang He.
\newblock Top-aware reinforcement learning based recommendation.
\newblock \emph{Neurocomputing}, 417:\penalty0 255--269, December
  2020{\natexlab{a}}.
\newblock ISSN 0925-2312.
\newblock \doi{10.1016/j.neucom.2020.07.057}.
\newblock URL
  \url{https://www.sciencedirect.com/science/article/pii/S0925231220311656}.
\newblock Accessed on 2021-03-09.

\bibitem[Liu et~al.(2020{\natexlab{b}})Liu, Tang, Li, Zhang, Ye, Chen, Guo,
  Zhang, and He]{liu_state_2020}
Feng Liu, Ruiming Tang, Xutao Li, Weinan Zhang, Yunming Ye, Haokun Chen,
  Huifeng Guo, Yuzhou Zhang, and Xiuqiang He.
\newblock State representation modeling for deep reinforcement learning based
  recommendation.
\newblock \emph{Knowledge-Based Systems}, 205:\penalty0 106170, October
  2020{\natexlab{b}}.
\newblock ISSN 0950-7051.
\newblock \doi{10.1016/j.knosys.2020.106170}.
\newblock URL
  \url{https://www.sciencedirect.com/science/article/pii/S095070512030407X}.
\newblock Accessed on 2021-03-09.

\bibitem[Liu et~al.(2020{\natexlab{c}})Liu, Guo, Li, Tang, Ye, and
  He]{liu_end--end_2020}
Feng Liu, Huifeng Guo, Xutao Li, Ruiming Tang, Yunming Ye, and Xiuqiang He.
\newblock End-to-{End} {Deep} {Reinforcement} {Learning} based {Recommendation}
  with {Supervised} {Embedding}.
\newblock In \emph{Proceedings of the 13th {International} {Conference} on
  {Web} {Search} and {Data} {Mining}}, {WSDM} '20, pages 384--392, New York,
  NY, USA, January 2020{\natexlab{c}}. Association for Computing Machinery.
\newblock ISBN 978-1-4503-6822-3.
\newblock \doi{10.1145/3336191.3371858}.
\newblock URL \url{https://doi.org/10.1145/3336191.3371858}.
\newblock Accessed on 2021-03-09.

\end{thebibliography}

\begin{appendices}

\chapter{Network Architectures}
\label{ap:architectures}

\begin{figure}
	\centering
	\includegraphics[width=0.7\textwidth]{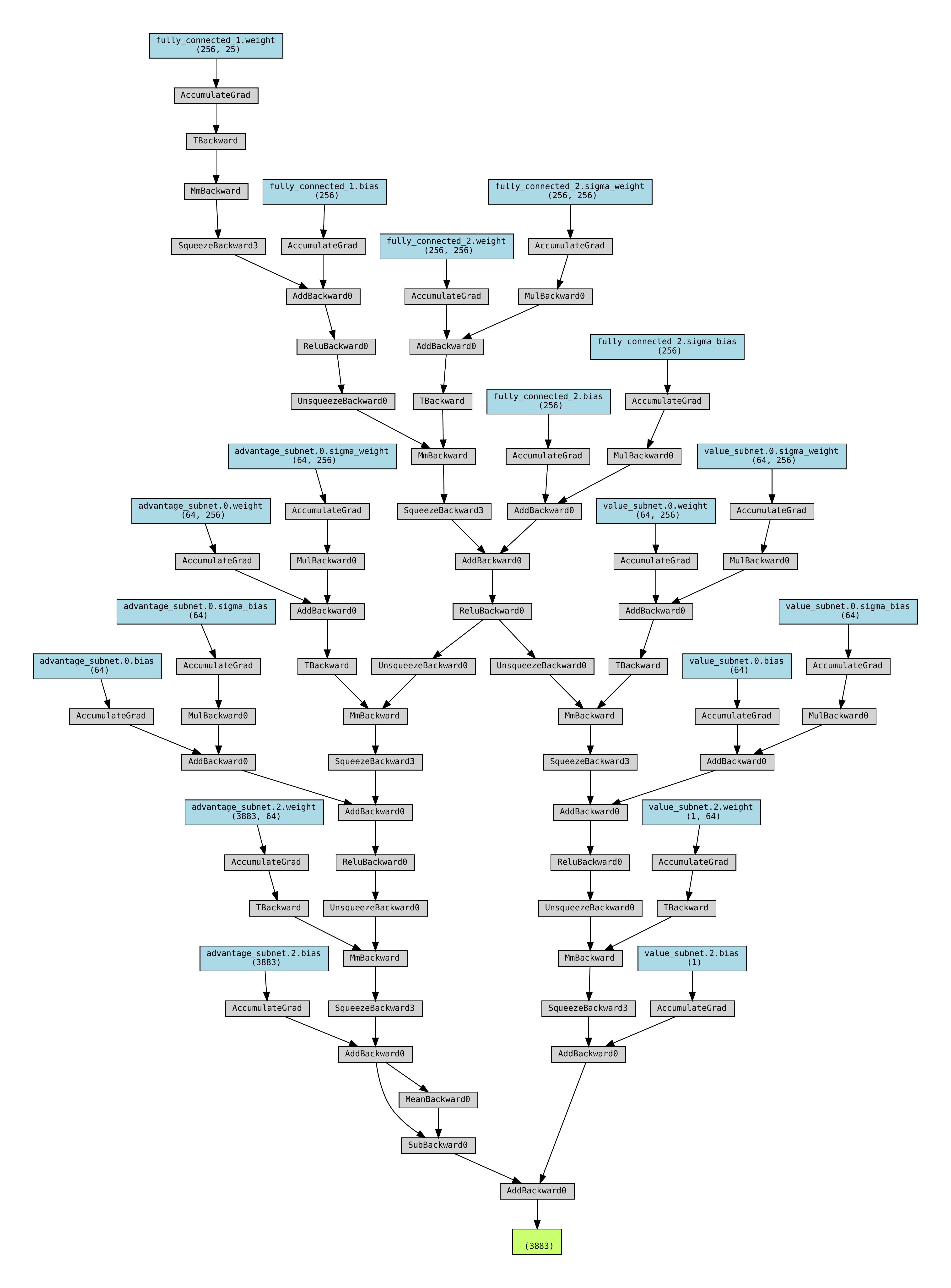}
	\caption{Dueling DQN gradient graph.}
	\label{fig:duelingqestimator}
\end{figure}

\begin{figure}
	\centering
	\includegraphics[width=0.7\textwidth]{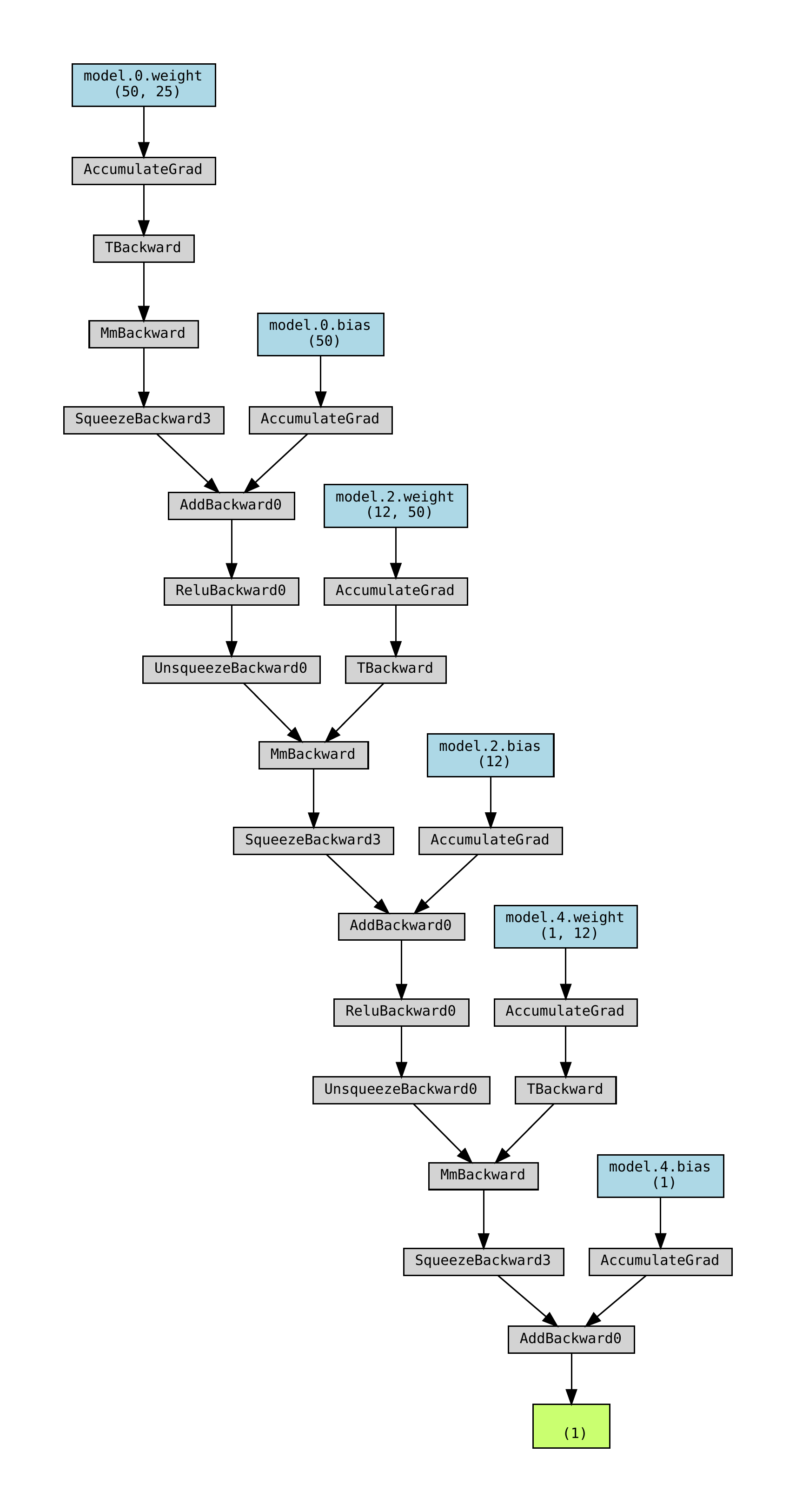}
	\caption{Value estimator gradient graph.}
	\label{fig:valueestimator}
\end{figure}

\begin{figure}
	\centering
	\includegraphics[width=0.7\textwidth]{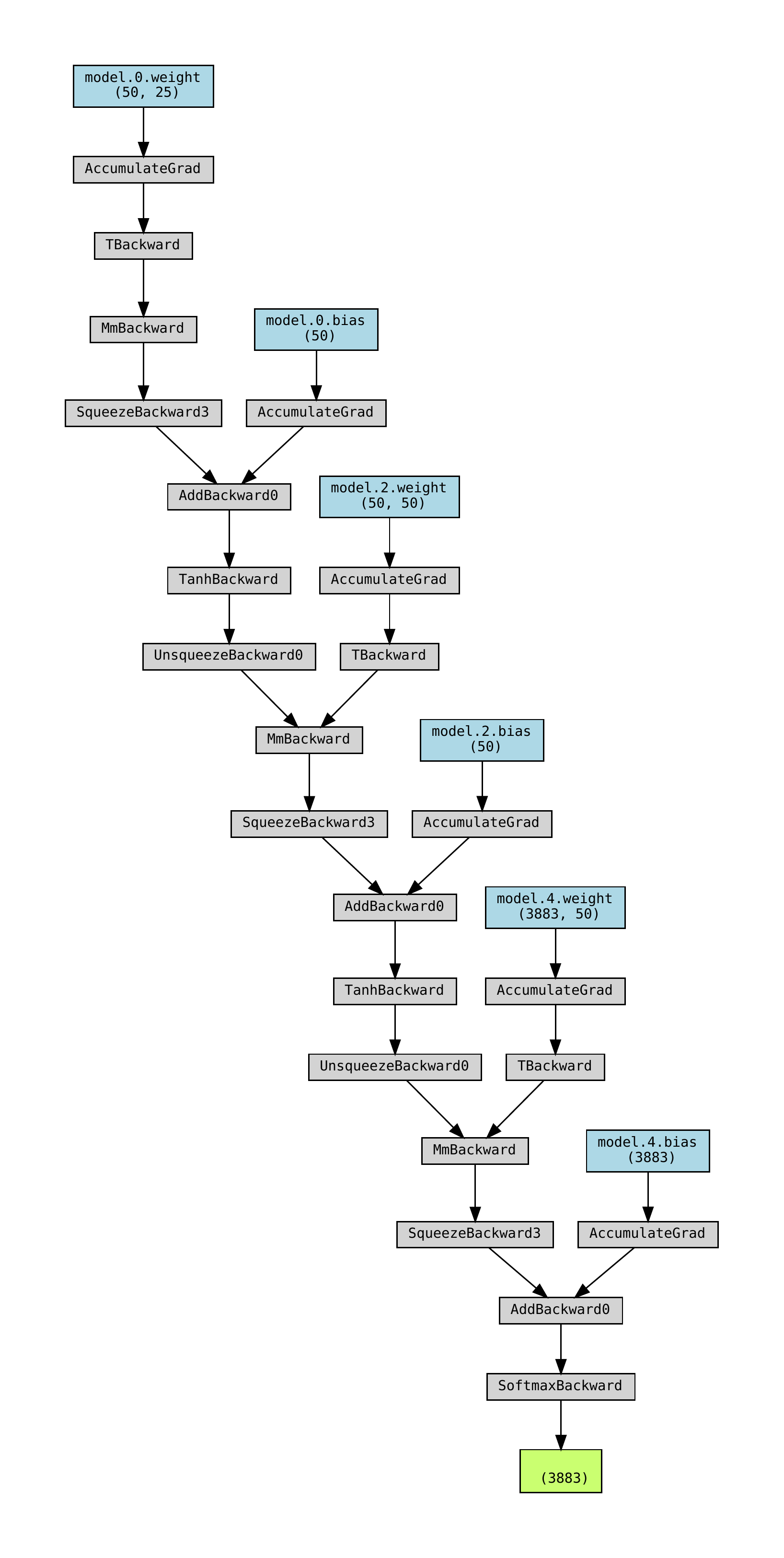}
	\caption{Policy estimator gradient graph.}
	\label{fig:policyestimator}
\end{figure}

\chapter{Hyperparameter Search Results}
\label{ap:paramsresults}

\begin{table}[!htbp]
\centering
\begin{tabular}{l}
\begin{filecontents*}{csv/hyperparams_ac.csv}
Agent,Parameter,Value,Seed 1,Seed 2,Seed 3
ActorCritic,critic hidden layers,[64 32],25.1849,23.7950,23.8909
ActorCritic,critic hidden layers,[128 64],26.6424,28.3380,28.6798
ActorCritic,critic hidden layers,[256 128],26.9030,20.4505,24.0356
ActorCritic,discount factor,0.9,32.5063,24.4924,22.0036
ActorCritic,discount factor,0.95,24.3296,24.5951,22.5244
ActorCritic,discount factor,0.99,26.2057,27.1427,24.5305
ActorCritic,critic learning rate,1e-05,21.1772,27.8518,22.9904
ActorCritic,critic learning rate,0.0001,23.8572,25.0933,23.0319
ActorCritic,critic learning rate,0.001,24.1299,22.4342,21.4234
Reinforce,hidden layers,[64 64],32.0943,31.8088,31.5308
Reinforce,hidden layers,[128 128],33.8032,33.6644,31.4792
Reinforce,hidden layers,[256 256],31.4428,32.6030,33.3391
Reinforce,discount factor,0.9,32.0863,30.8048,31.5693
Reinforce,discount factor,0.95,35.0739,31.8631,30.0414
Reinforce,discount factor,0.99,31.0604,31.3833,29.9745
Reinforce,learning rate,1e-05,33.4242,32.1575,31.0293
Reinforce,learning rate,0.0001,33.5311,32.7911,32.7734
Reinforce,learning rate,0.001,32.9169,32.7536,33.8805
\end{filecontents*}
\csvautotabular{csv/hyperparams_ac.csv}
\end{tabular}
\caption{Actor-Critic and REINFORCE hyperparameter search results.}
\label{table:hyperparams_ac}
\end{table}

\begin{table}
\centering
\begin{tabular}{l}
\begin{filecontents*}{csv/hyperparams_dqn.csv}
Agent,Parameter,Value,Seed 1,Seed 2,Seed 3
DuelingDQN,network update freq.,3,30.4253,34.0480,39.1959
DuelingDQN,network update freq.,5,33.2199,28.9370,29.6506
DuelingDQN,network update freq.,7,26.3449,31.2439,37.4598
DuelingDQN,network sync freq.,100,31.7379,21.0853,29.4332
DuelingDQN,network sync freq.,200,31.3731,29.3101,27.8519
DuelingDQN,network sync freq.,300,29.8941,31.2043,33.5061
DuelingDQN,priority importance,0.4,30.5110,31.8823,36.6239
DuelingDQN,priority importance,0.6,27.9894,30.5443,33.4694
DuelingDQN,priority importance,0.8,29.5587,29.5457,39.1650
DuelingDQN,priority weight growth,0.0001,29.8128,29.3818,35.5804
DuelingDQN,priority weight growth,0.001,31.8397,26.3807,35.7185
DuelingDQN,priority weight growth,0.01,30.0311,32.2882,33.9274
DuelingDQN,buffer size,5000,28.9867,30.3517,32.8997
DuelingDQN,buffer size,10000,34.8314,30.9879,31.6142
DuelingDQN,buffer burn in,500,29.2160,29.6616,29.5514
DuelingDQN,buffer burn in,1000,30.7920,38.3978,37.7760
DuelingDQN,buffer burn in,1500,30.5391,29.8369,30.7158
DuelingDQN,batch size,16,27.9476,30.5927,31.7432
DuelingDQN,batch size,32,33.9847,34.2276,32.9230
DuelingDQN,batch size,48,28.9723,33.2276,38.9923
DuelingDQN,noise sigma,0.1,30.4463,29.7764,33.7494
DuelingDQN,noise sigma,0.017,33.6226,31.9494,33.0891
DuelingDQN,noise sigma,0.25,28.7930,25.2711,27.8358
DuelingDQN,hidden layers,[128 128 32 32],30.6985,33.7101,36.7923
DuelingDQN,hidden layers,[256 256 64 64],31.1062,33.7865,32.8018
DuelingDQN,hidden layers,[512 512 128 128],34.3929,40.4306,41.7473
\end{filecontents*}
\csvautotabular{csv/hyperparams_dqn.csv}
\end{tabular}
\caption{Dueling DQN hyperparameter search results.}
\label{table:hyperparams_dqn}
\end{table}

\end{appendices}

\end{document}